\documentclass[pdflatex,sn-mathphys-ay]{sn-jnl}

\usepackage{graphicx}%
\usepackage{multirow}%
\usepackage{amsmath,amssymb,amsfonts}%
\usepackage{amsthm}%
\usepackage{mathrsfs}%
\usepackage[title]{appendix}%
\usepackage{xcolor}%
\usepackage{textcomp}%
\usepackage{manyfoot}%
\usepackage{booktabs}%
\usepackage{algorithm}%
\usepackage{algorithmicx}%
\usepackage{algpseudocode}%
\usepackage{listings}%

\theoremstyle{thmstyleone}%
\theoremstyle{thmstyletwo}%

\theoremstyle{thmstylethree}%

\newcommand{\ex}{\mathbf{e}_{x}}
\newcommand{\ey}{\mathbf{e}_{y}}
\newcommand{\ez}{\mathbf{e}_{z}}

\newcommand{\exprim}{\mathbf{e}_{X'}}
\newcommand{\eyprim}{\mathbf{e}_{Y'}}
\newcommand{\ezprim}{\mathbf{e}_{Z'}}
\newcommand{\sign}{\,{\rm sign}\,}
\newcommand{\atanh}{\,{\rm atanh}\,}
\newcommand{\cotan}{{\rm cotan}\,}

\newcommand{\psiref}{\Psi_{\rm ref.}}

\newcommand{\apss}{Astrophys. Space Sci/}
\newcommand{\planss}{Planet. Space Sci.}

\newcommand{\aj}{Astron. Journal}
\newcommand{\aap}{Astron. \& Astrophys.}
\newcommand{\mnras}{Monthly Notices of the Royal Astronomical Society}
\newcommand{\na}{New Astron.}
\newcommand{\icarus}{Icarus}
\newcommand{\psj}{Planet. Sci. J.}
\newcommand{\maps}{Meteorit. \& Planet. Sci}

\begin{document}

\title[Gravity from surface triangulation]{Gravity from surface triangulation: convergence acceleration with nested grids}

\author{\fnm{Hur\'e}\sur{Jean-Marc}}\email{jean-marc.hure@u-bordeaux.fr}



\affil{\orgdiv{University of Bordeaux, CNRS, Laboratoire d'Astrophysique de Bordeaux, UMR 5804}, \orgaddress{\city{Pessac}, \postcode{33600}, \country{France}}}

\abstract{The determination of the gravitational potential by the polyhedral method is revisited in the case where the surface of a body is composed of triangular facets. Based upon six test-shapes of astrophysical interest (sphere, spheroid, triaxial, lemon-shape, dumbell and torus) projected on nested grids, we verify that the convergence toward reference values is second-order in the step size of the grid, inside the body, at the surface and outside. We then show that the accuracy or computing time can be drastically enhanced by implementing the Repeated Richardson Extrapolation. This technique is especially efficient when the body's surface is smooth enough, and is therefore well adapted to the theory of figures (single and multi-layer fluids) and to dynamical studies (test-particle and mutual interactions), which require a large number of field evaluations. For real objects like asteroids that have very irregular terrains at small scales, the gain is modest. In that context, we estimate the discretization level beyond which the typical error in potential values due to altimetric uncertainties dominates over the contribution of sub-grid cavities and bumps. For bodies close to spherical, the criterion reads $T \gtrsim \frac{64 D}{3 \lambda},$  where  $D$ is the diameter of the body, $\lambda$ the typical shape error and $T$ the number of triangular facets involved. The case of 433 Eros is considered as an example.}




\maketitle
\section{Introduction and motivations}\label{sec1}

The polyhedral decomposition is a well-proven technique for estimating the gravity field of certain celestial bodies with complex shape \citep{werner94}. Homogeneous objects whose surface is entirely made of planar facets (bounded by straight edges) are actually among rare cases for which the expressions for the Newtonian potential and accelerations are mathematically exact, expressible in closed form from Cartesian coordinates and singularity-free \citep{p74,wal76,c12}. Even if real bodies are covered with cavities and bumps and exhibit mass-density variations, this approach is of immense interest for analyzing the structure, composition and surface of asteroids, satellites and dwarf planets from topological data \citep[e.g.][]{hpt09,rkbepg14,durso14bis,ul24,yb15,bal20}, the mutual interaction of aggregates and rubber piles, their growth, stability and disruption  \citep{k03,k04,ka05,mbr04,fs06,ft20,hwpc23,lsz25}, and for investigating the trajectory of test-particles or probes orbiting around \citep{smy02,yl18,jx19,kvt21,pp22,fsist24,pbms24}.

In terms of computational cost, the Polyhedron Gravity Method (PGM) is comparable to an elementary multi-body code \citep{binneytremaine87}. A closed surface discretized into $N \times N$ nodes contains $2N^2$ triangles (and the same amount of tetrahedra) typically, and the time $\tau(N)$ required to compute the potential or acceleration at a given point in space is proportional to $N^2$, and it is to $N^4$ for a full coverage (ie. all nodes visited). This becomes prohibitive for large grids, especially when computing the trajectory of test-particles. On the other hand, the accuracy, traditionnally measured through the error $E(N)$, increases with $N$ (for a second-order scheme, this is $E(N) \propto 1/N^2$). But accuracy is also limited by the arithmetic of computers \citep[e.g.,][]{opac-b1132370}. Raising the resolution is therefore at the expense of the computing time, and one generally seeks for a compromise between accuracy and rapidity of algorithms. In these conditions, it appears that the only way to reduce the computional cost $\tau$ without impacting the quality of outputs is parallel and/or GPU-based processing \citep{wz23,sbgig24}, which enables to reach the lower limit $\tau \propto N^2$.

This paper deals with the Tetrahedron Gravity Method (TGM), a subclass of the PGM, where all primitive facets of the body's surface are triangles. We investigate the degree of convergence of potential values in the vicinity of the body as the number of facets increases. We show that a given accuracy level can be reached by implementing a classical numerical recipe of convergence acceleration, which enables to reduce the actual grid size ($N \rightarrow N' < N$), thereby decreasing $\tau$ again, possibly by orders of magnitude depending on the smoothness of the surface in question. In Sect. \ref{sec:solution}, we shortly revisit the TGM by deriving a collection of compact formula established in local coordinates. In Sect. \ref{sec:test}, we test the accuracy of the method by considering six test-shapes of astrophysical interest (simply and multiply connected domains). Comparisons with references (e.g., sphere, triaxial) is essential. To our knowledge, this has never been published yet. In particular, by using nested grids (required for convergence acceleration), we verify first that that potential values are second-order accurate in the mesh spacing for any evaluation point. The conditions are therefore ideal to apply a Repeated Richardson Extrapolation (RRE), which enables to generate much better values without significant computational effort \citep{frr07}. This is the aim of Sect. \ref{sec:rre}. As demonstrated, the performances of the RRE technique are optimal if the surface in consideration is smooth enough, with no altimetric variations below a certain length scale. We then discuss the efficiency of the method in a real case for asteroid 433 Eros by using shape data from the NASA Data Planetary System \citep{gaskell21}. As expected and tested, the RRE technique is less usefull, due to terrain irreguarities at the smallest scales. The impact of shape uncertainty in potential values is discussed \citep[e.g.][]{bpm20}, and compared to the contribution of small-scale cavities and bumps. Quantitative gains in terms of computing time and error reduction are outlined in Sect. \ref{sec:gain}. A few perspectives are listed in the conclusion.

\section{Statement of the problem and exact solution}
\label{sec:solution}

The Newtonian potential at point P$(\mathbf{r})$ in space of a homogeneous body is given by the fundamental expression \citep[e.g.,][]{kellogg29,durand64}
\begin{flalign}
\Psi( \mathbf r)= -G\rho \iiint_{\cal V}\frac{dV'}{|\mathbf{r}-\mathbf{r'}|},
\label{eq:tripleint}
\end{flalign}
where $\rho$ is the mass density and ${\cal V}$ is to the total volume occupied by matter. The presence of sharp edges and vertices in the geometry (like for a cube for instance) does not produce any kind of singularities, for the potential itself, and its first and second spatial derivatives according to the Poisson equation \citep{durand64,wal76,cha12,durso14}. By applying the divergence Gauss-Ostrogradsky theorem, we get the equivalent form
\begin{flalign}
\Psi(\mathbf{r})= + \frac{1}{2}G \rho \iint_{\cal S}\mathbf{u}(\mathbf{r}|\mathbf{r}') \cdot \mathbf{n}(\mathbf{r}') dS',
\label{eq:doubleint}
\end{flalign}
where ${\cal S}$ is the total surface of the body associated with the volume ${\cal V}$, $\mathbf{n}$ is the local unit vector, normal to the surface oriented outwards, and
\begin{flalign}
  \mathbf{u}(\mathbf{r}|\mathbf{r}') = \frac{\mathbf{r}-\mathbf{r'}}{|\mathbf{r}-\mathbf{r'}|}.
\end{flalign}
The main advantage of this double integral is that the hyperbolic singularity in the integrand of Eq. \eqref{eq:tripleint} is removed as $\mathbf{u}(\mathbf{r}|\mathbf{r}') \cdot \mathbf{n}$ is just a cosine. This integral can be converted into a line integral from the curl theorem due to Stokes, but this does not necessarily renders the analytical treatment much easier \citep[e.g.,][]{werner94,z09,tsoulis77,c15,mm18}.

\subsection{Surface triangulation}

\begin{figure}
       \centering
       \includegraphics[width=8.5cm,trim={8cm 4cm 1cm 3cm},clip]{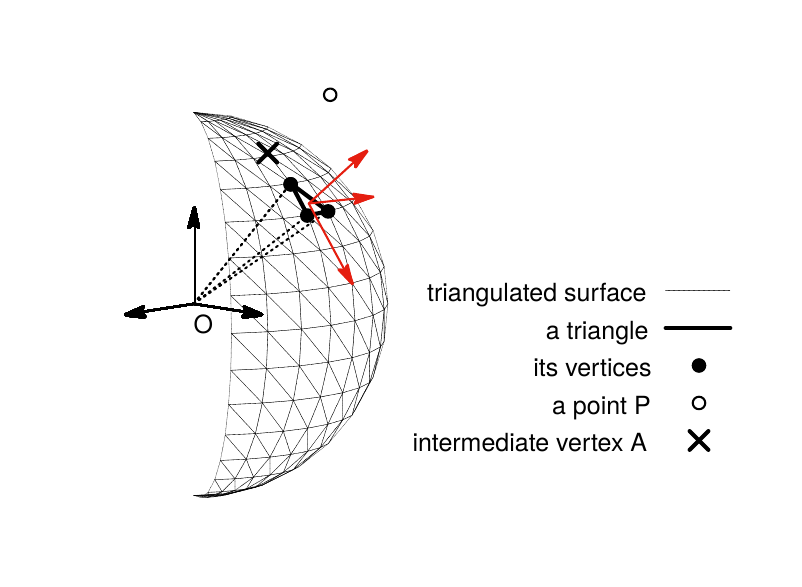}
       \caption{Part of a sphere triangulated with $960$ triangles obtained with $N=2M=32$ equally-spaced nodes along the azimuth $\phi$ and $M$ equally-spaced nodes in colatitude $\theta$. Also shown are a particular triangle ({\it bold line}), the local basis vectors ({\it red arrows}), a point P$(x,y,z)$ of space where the potential $\Psi$ is evaluated, the intermediate vertex A defined as the projection of point P in the plane of the actual triangle ({\it black cross}), and the basis vectors of the global reference frame ({\it black arrows})}
       \label{fig:sphere.pdf}
\end{figure}

In practice, the double integral in Eq. \eqref{eq:doubleint} is calculated by discretization of the surface ${\cal S}$, in the form of connected nodes V$_i(\mathbf{r}_i)$. The most basic bidimensionnal pattern in flat geometry being the triangle, we assume here that the nodes can be associated by triplets $\{$V$_i,$V$_j,$V$_k\} \equiv {\cal T}_{ijk}$ forming non-intersecting triangles ${\cal T}_{ijk}$, sharing edges and vertices. Regarding Eq. \eqref{eq:tripleint}, this is equivalent to an assembly of tetrahedra $\{$O$,$V$_i,$V$_j,$V$_k\}$ where point O is, for instance, the center of mass. Figure~\ref{fig:sphere.pdf} shows an example of triangulated sphere with unit radius (see Sect.~\ref{subsec:test3geometries} for the definition for the vertices). In these conditions, Eq.\eqref{eq:doubleint} becomes
\begin{flalign}
\Psi(\mathbf{r})= \sum_{\text{all triangles}}{\delta \Psi_{ijk}(\mathbf{r})},
\label{eq:psitotal}
\end{flalign}
where $\delta \Psi_{ijk}$ is the contribution of triangle ${\cal T}_{ijk}$, namely
\begin{flalign}
\delta \Psi_{ijk} = + \frac{1}{2}G \rho \iint_{\text{triangle }{\cal T}_{ijk}}\mathbf{u}(\mathbf{r}|\mathbf{r}') \cdot \mathbf{n} dS'.
\label{eq:psiijk}
\end{flalign}

At this level, the labels $i$, $j$ and $k$ are arbitrary but must refer to a unique, non-ambiguous numbering of nodes. We do not go into the details of mesh generation, which is well-documented \citep[e.g.,][]{pj21}.

\subsection{Local coordinate system and intermediate vertex A}

\begin{figure}
       \centering
       \includegraphics[width=8.5cm,trim={1.cm 41cm 0.cm 0.cm},clip]{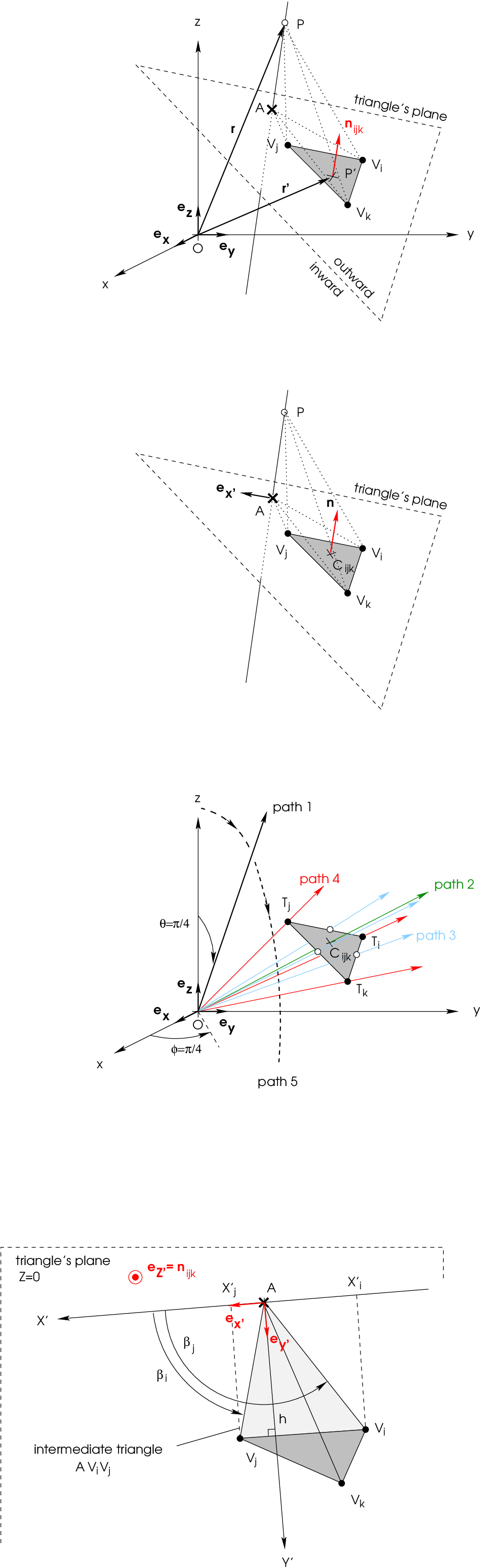}
       \caption{A triangle ${\cal T}_{ijk}=\{$V$_i$,V$_j$,V$_k\}$ as part of the triangulated surface ${\cal S}$ of the body, its plane and the normal vector $\mathbf{n}_{ijk}$, oriented outwards. Point A is the projection of P in this plane following its normal $\mathbf{n}_{ijk}$, leaving $3$ intermediate triangles AV$_i$V$_j$, AV$_j$V$_k$ and AV$_k$V$_i$; see Fig.~\ref{fig:configV2.eps}}
       \label{fig:config.eps}
\end{figure}

\begin{figure}
       \centering
       \includegraphics[width=8.5cm,trim={0.cm 0.cm 1.cm 41.cm},clip]{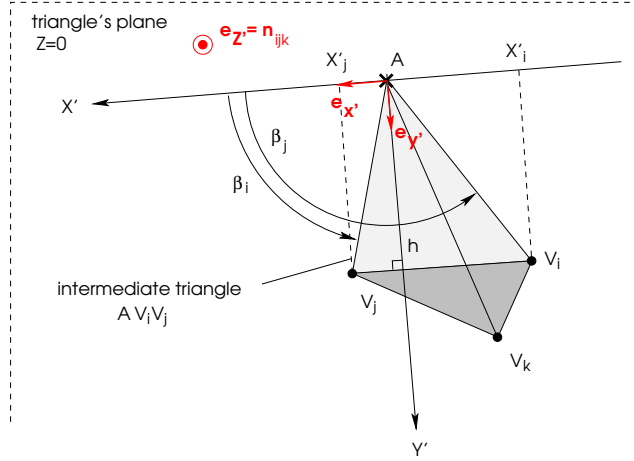}
       \caption{Local Cartesian coordinates $(X',Y',Z')$ in the plane of the triangle ${\cal T}_{ijk}=\{$V$_i$,V$_j$,V$_k\}$ ({\it dark gray}) and local basis is $(\exprim,\eyprim,\ezprim)$, with $\ezprim \equiv \mathbf{n}_{ijk}$; see Fig. \ref{fig:config.eps}. The height of the intermediate triangle AV$_i$V$_j$ ({\it light gray}) is $h$}
       \label{fig:configV2.eps}
\end{figure}

Inspired by \cite{Stirling_2017} and \cite{amaral19}, we calculate $\delta \Psi_{ijk}$ by using a local coordinate system, which is sometimes very efficient for establishing a compact formula and avoiding explicit singularities \citep{tisserand91,p74,wal76,lassblitzer83}. For a given triangle ${\cal T}_{ijk}$, the local system is defined by the triangle's plane and its unit normal vector $\mathbf{n}_{ijk}$ oriented outwards, as depicted in Fig.~\ref{fig:config.eps}. In this plane, the coordinates are denoted $(X',Y',Z')$. The local basis $(\exprim,\eyprim,\ezprim)_{ijk}$ is formed by a few trivial vectorial manipulations. We first define the unit vector $\mathbf{t}_{ij}$ such that $||\mathbf{V}_i\mathbf{V}_j|| \mathbf{t}_{ij} = \mathbf{V}_i\mathbf{V}_j$ and set $\exprim \equiv \mathbf{t}_{ij}$. Next, from $\cos \alpha = \mathbf{t}_{ij} \cdot \mathbf{t}_{ik}$, we have
\begin{flalign}
  \ezprim \equiv \frac{\mathbf{t}_{ij} \times \mathbf{t}_{jk}}{\sin \alpha} \equiv \mathbf{n}_{ijk},
\end{flalign}
where $||\mathbf{V}_i\mathbf{V}_k|| \mathbf{t}_{ik} = \mathbf{V}_i\mathbf{V}_k$ and $||\mathbf{V}_j\mathbf{V}_k|| \mathbf{t}_{jk} = \mathbf{V}_j\mathbf{V}_k$. The last vector of the local basis is $\eyprim \equiv \ezprim \times \exprim$. In the local frame, the Cartesian coordinates are denoted $(X',Y',Z)$, with the consequence that the point P'$(X',Y',0)$ belongs to the triangle's plane. Finally, we consider the point A as the projection of the evaluation point P onto the plane of the triangle (see below); see Figs.~\ref{fig:sphere.pdf} and \ref{fig:config.eps}. This point is also set as the origin of the local coordinate system, i.e., A$(0,0,0)$. Three intermediate, coplanar triangles AV$_i$V$_j$, AV$_j$V$_k$ and AV$_k$V$_i$ are then formed, as shown in Fig.~\ref{fig:configV2.eps}. Note that, as done for the unit vector $\mathbf{n}$, points A is specific to each triangle and must be formally denoted A$_{ijk}$. Similarly, P$(x,y,z)$ is unique but its coordinates in the local frame, i.e., $(X',Y',Z')$, are specific to each triangle, and so we should write P$_{ijk}$ and the local altidude is $Z' \equiv Z'_{ijk}$. We omit these extra notations in the following for the comfort of reading.

\subsection{Double integration over the 3 intermediate triangles}

In the local coordinate system, we have P$(0,0,Z')$ and it follows that
\begin{flalign}
  \mathbf{r}-\mathbf{r'}=-X'\exprim-Y'\eyprim+Z' \ezprim.
\end{flalign}
Unless P lies in the triangle's plane, $Z'$ is nonzero (see below). As $dS'=dX'dY'$, we have
\begin{flalign}
\frac{dS'}{|\mathbf{r}-\mathbf{r'}|}=\frac{dX'dY'}{\sqrt{{X'}^2 + {Y'}^2+{Z'}^2}},
\end{flalign}
and so the integrand in Eq. \eqref{eq:psiijk} is
\begin{flalign}
\mathbf{u}(\mathbf{r}|\mathbf{r}') \cdot \mathbf{n}_{ijk} dS' = Z' \frac{dX'dY'}{\sqrt{{X'}^2 + {Y'}^2+{Z'}^2}}.
\label{eq:dsprim}
\end{flalign}
That expression is attractive because it is fully integrable over $X'$ and $Y'$ by analytical means. In this purpose, it is interesting to split the integral in Eq. \eqref{eq:psiijk} into three integrals running over the area of the $3$ intermediate triangles AV$_i$V$_j$, AV$_j$V$_k$ and AV$_k$V$_i$; see again Fig.~\ref{fig:configV2.eps}. For the first intermediate triangle AV$_i$V$_j$, we integrate first along the $X'$ axis, and we find
\begin{flalign}
&\int_{X'_i}^{X'_j} {\frac{dX'}{\sqrt{{X'}^2 + {Y'}^2+{Z'}^2}}} = \left[\atanh \frac{X'}{\sqrt{{X'}^2 + {Y'}^2+{Z'}^2}}\right]_{X'_i}^{X'_j},
\label{eq:intxprim_dsprim}
\end{flalign}
where the lower and upper bounds are $X'_i(Y')=Y' \cotan \beta_i$ and $X'_j(Y')=Y' \cotan \beta_j$, respectively, and the $\beta$'s are measured with respect to $\exprim$, i.e., $\beta_i=(\exprim,\mathbf{AV}_i)$, with a similar definition for $\beta_j$. These angles must be signed, automatically ensuring the right sign of the integral. The next integration runs over the variable $Y'$. For the lower-bound of Eq. \eqref{eq:intxprim_dsprim}, we have
\begin{flalign}
\int_{0}^{h} {\atanh \frac{ \cotan \beta_i Y'}{\sqrt{(1+ \cotan^2 \beta_i){Y'}^2+{Z'}^2}}  dY' } \equiv I(\cotan \beta_i,Z'; h),
  \label{eq:i}
\end{flalign}
where $I(a,b;x)$ is fully analytical (see the Appendix \ref{app:i}), and $h$ denotes the height of the triangle AV$_i$V$_j$. This is positive quantity  defined by $h=$AV$_i \sin \beta_i= $AV$_j\sin \beta_j$. The formula is similar for the upper-bound of Eq.\eqref{eq:intxprim_dsprim}. It follows that the full integration of Eq. \eqref{eq:dsprim} over $X'$ and $Y'$ for the first intermediate triangle AV$_i$V$_j$ is
\begin{flalign}
  \label{eq:intAViVj}
  I_{\text{AV}_i\text{V}_j} & =  \iint_{\text{triangle AV}_i\text{V}_j} \frac{dX'dY'}{\sqrt{{X'}^2 + {Y'}^2+{Z'}^2}}\\
  &= I(\cotan \beta_j,Z';h)-I(\cotan \beta_i,Z';h),\nonumber
\end{flalign}
which depends uniquely on the height $h$ of the intermediate triangle, on the two angles $\beta_i$ and $\beta_j$, and on the relative position of point P, i.e., $Z'$. It can be verified that Eq. \eqref{eq:intAViVj} is valid and finite for any point A relative to the vector $\mathbf{V}_i\mathbf{V}_j$, including cases where A is at a vertex. Note that the two angles $\beta_i$ and $\beta_j$ need to be renamed $\beta_{ij,i}$ and $\beta_{ij,j}$, respectively, because these refer to the triangle $\text{AV}_i\text{V}_j$ specifically, and there are three intermediate triangles in total. In a similar manner, the height $h$ becomes $h_{ij}$.

We repeat the calculation for the other two intermediate triangles AV$_j$V$_k$ and AV$_k$V$_i$ with the same local coordinate system, leading to
\begin{flalign}
  I_{\text{AV}_j\text{V}_k} =I(\cotan \beta_{jk,k},Z';h_{jk})-I(\cotan \beta_{jk,j},Z';h_{jk}),
\end{flalign}
and
\begin{flalign}
  I_{\text{AV}_k\text{V}_i} = I(\cotan \beta_{ki,k},Z';h_{ki})-I(\cotan \beta_{ki,i},Z';h_{jk}).
\end{flalign}
These formula are not new \citep{p74,wal76,werner94,hk96,c12,c15}, and can take different forms \citep{tsoulis77}.

\subsection{Contribution of a triangle}

The integral over the triangular domain ${\cal T}_{ijk}$ is a combination of the integrals over the $3$ intermediate triangles, namely $I_{\text{AV}_i\text{V}_j}$, $I_{\text{AV}_j\text{V}_k}$ and $I_{\text{AV}_k\text{V}_i}$. The result is therefore of the form
\begin{flalign}
  \label{eq:integijk}
  \iint_{\text{triangle } {\cal T}_{ijk}}{\frac{dS'}{|\mathbf{r}-\mathbf{r'}|}} = s_{ij} I_{\text{AV}_i\text{V}_j} +   s_{jk} I_{\text{AV}_j\text{V}_k} +  s_{ki} I_{\text{AV}_k\text{V}_i},
\end{flalign}
where the coefficients $s \in \{-1,0,+1\}$. t can be verified that
\begin{flalign}
s_{ij} = \sign [ \ezprim \cdot (\mathbf{AV}_i \times \mathbf{AV}_j) ],
\end{flalign}
 and with a similar definition for $s_{jk}$ and $s_{ki}$. Then Eq. \eqref{eq:psiijk} becomes
\begin{flalign}
  \delta \Psi_{ijk} = +\frac{1}{2}G \rho Z' \left(s_{ij} I_{\text{AV}_i\text{V}_j} + s_{jk} I_{\text{AV}_j\text{V}_k} +  s_{ki} I_{\text{AV}_k\text{V}_i}\right).
  \label{eq:deltapsiijk}
\end{flalign}

Note that we have:
\begin{itemize}
\item $s_{ij}=0$, when point A stands on the line ($\text{V}_i\text{V}_j)$, and then $s_{jk}=s_{ki}=+1$. The remark is similar for $s_{jk}$ and $s_{ki}$.
\item $s_{ij}=s_{jk}=0$, when point A is at the vertex V$_j$, and then $s_{ki}=+1$. The remark is similar for the next two vertices.
\item $s_{ij}=s_{jk}=s_{jk}=+1$, when point A stands inside the triangle $\text{triangle T}_i\text{V}_j\text{V}_k$.
\end{itemize}

\subsection{Summing all contributions}

The above procedure, dedicated to a single triangle, is to be repeated for all triangles composing the closed surface, and there is no difficulty in programming the method. A special attention must, however, be paid to treat situations where the height $h \rightarrow 0$ and $\beta \rightarrow \pm \frac{\pi}{2}$ to avoid spurious, diverging values. For each triangle in the sample, one must determine the intermediate vertex $A$, three heights $h$ and six angles $\beta$. As the surface is generally defined in a fixed, non-local frame, it is necessary to connect each local coordinate system to the general coordinate system $($O$,\ex,\ey,\ez)$. The conversion formula are given in the Appendix \ref {app:avertex}. See the Appendix \ref{app:comments} for a few comments.

\section{Testing the direct method}
\label{sec:test}

\subsection{Three test-geometries with known reference}
\label{subsec:test3geometries}

The potential is known in closed form for the cube \citep[e.g.,][]{cha12} and its surface is decomposable into $12$ triangles. The cube is therefore a perfect reference body for benchmarking the performance of the potential computation outlined above. More interesting tests concern systems with curved surfaces, and we begin with the following shapes: 
\begin{itemize}
  \setlength{\itemsep}{+0pt}
\item a sphere with radius $a$, as illustrated in Fig.~\ref{fig:sphere.pdf},
\item a spheroid (i.e., an ellipsoid of revolution) with polar radius $c$ and equatorial radius $a$,
\item a triaxial ellipsoid with $a$ as the semi-major axis, and $b$ and $c$ as the semi-minor axis.
\end{itemize}

\begin{figure}
       \centering
       \includegraphics[width=8.5cm,trim={7cm 4cm 1cm 3cm},clip]{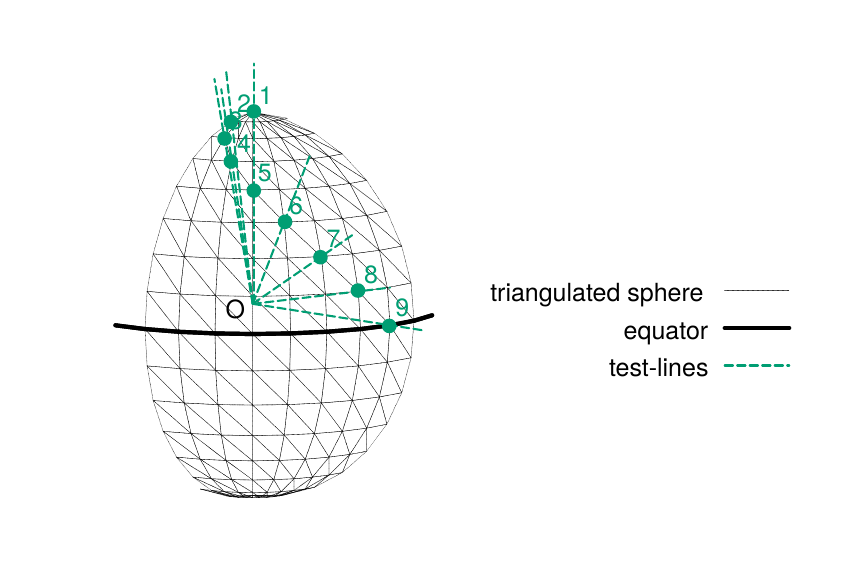}
       \caption{Same caption as for Fig.~\ref{fig:sphere.pdf}, but showing the test-lines defined by Eq. \eqref{eq:kspace} along which the potential is compared with references values}
       \label{fig:sphere_testlines.pdf}
\end{figure}

\begin{figure}
       \centering
       \includegraphics[height=7.5cm,trim={6.8cm 0.cm 6.3cm 1.1cm},clip]{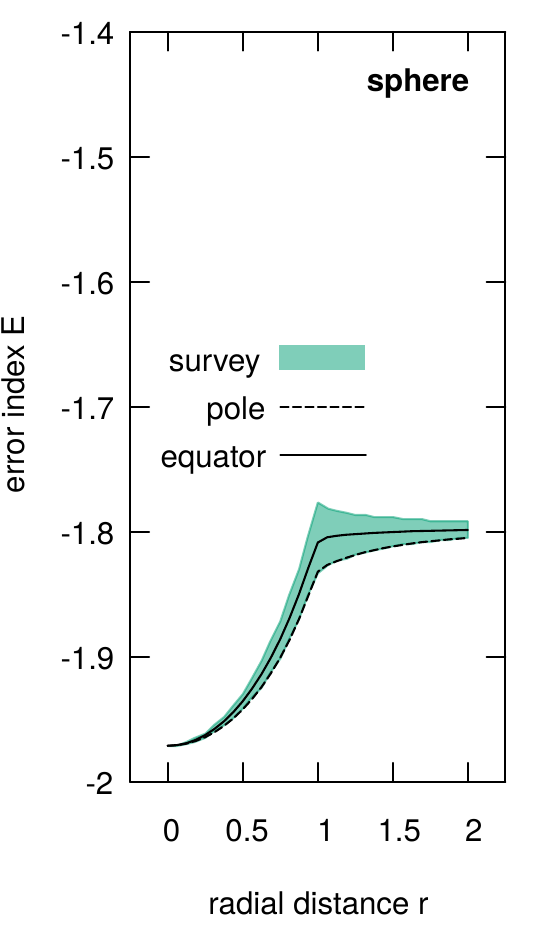}\includegraphics[height=7.5cm,trim={9cm 0.cm 6.3cm 1.1cm},clip]{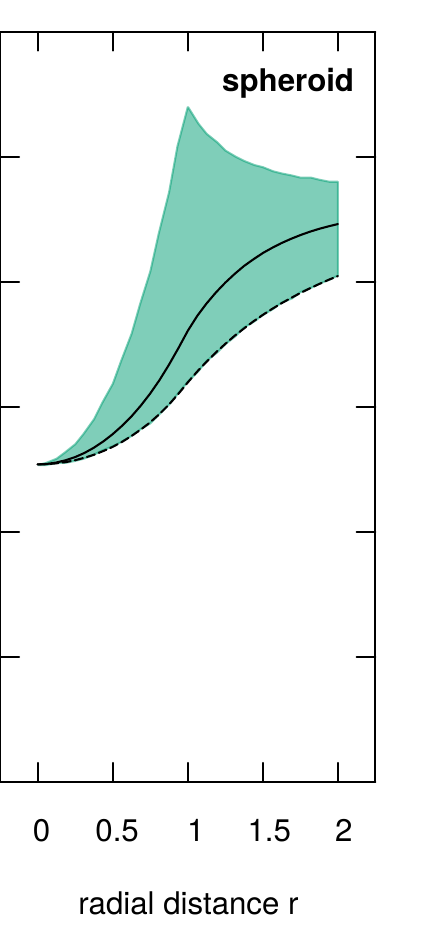}\includegraphics[height=7.5cm,trim={9cm 0.cm 7.2cm 1.1cm},clip]{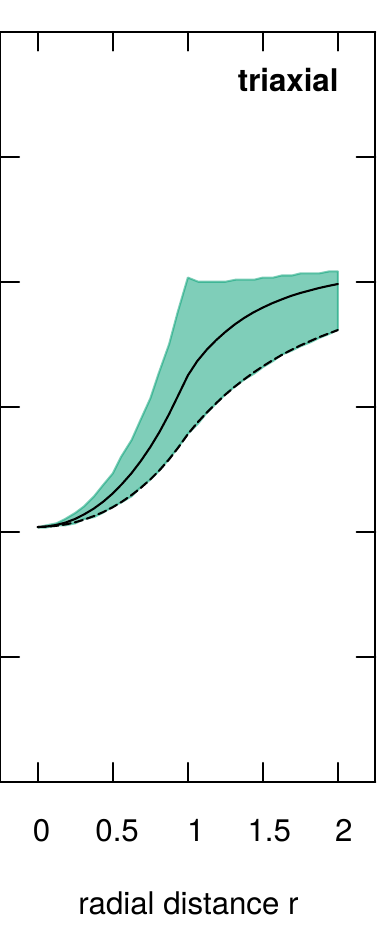}
       \caption{Error index $E$ for the gravitational potential computed from surface triangulation by direct summation, i.e., from Eq. \eqref{eq:psitotal}, for the sphere with $a=1$ ({\it left}), for the spheroid with $a=1$ and $b=3a/4$ ({\it middle}) and for the triaxial ellipsoid with $a=1$, $c=a/2$ and $b=3a/4$ ({\it right}) along the nine test-lines defined by Eq. \eqref{eq:kspace}, with $0 \le k \le K$ and $K=32$. The surface triangulation is performed with $N=2M=32$; see Fig.~\ref{fig:sphere_testlines.pdf} for the sphere}
       \label{fig:test_sphere.pdf}
\end{figure}

In the purpose of the tests, we use spherical coordinates $(r,\theta,\phi)$ to generate $M \times N$  angles according to
\begin{flalign}
  \begin{cases}
\theta_{j}=\pi\frac{j}{M} \; \text{(polar angle)} , \qquad j \in [0,M],\\
\phi_{i}=2\pi\frac{i}{N} \; \text{(azimuth)} , \qquad i \in [0,N].
  \end{cases}
  \label{eq:sphmesh}
\end{flalign}
At each vertex V$_{ij}(\theta_j,\phi_i)$, we calculate the radius $r_{ij} \equiv r(\phi_{i},\theta_{j})$ to model the sphere, the spheroid and the triaxial. In these conditions\footnote{For each quadruplet $\{$V$_{i,j},$V$_{i+1,j},$V$_{i,j+1},$V$_{i+1,j+1}\}$ with $i \in [0,N-1]$ and $j \in [1,M-2]$, we build two triangles with vertices at $\{$V$_{i,j},$V$_{i,j+1},$V$_{i+1,j+1}\}$ for the first one, and $\{$V$_{i,j},$V$_{i+1,j+1},$V$_{i+1,j}\}$ for the second one. For $j \in \{0,M-1\}$ which correspond to the poles, there are only two triangles, namely $\{$V$_{i,0},$V$_{i,1},$V$_{i+1,1}\}$ and $\{$V$_{i,M-1},$V$_{i,M},$V$_{i+1,M-1}\}$ for any given index $i \in [0,N-1]$; again, see Fig.~\ref{fig:sphere.pdf}.} there are $T=2N(M-1)$ triangles in total. We take $N=2^\ell$ where $\ell$ is the discretization level, which enables to easily generate nested grids (see below). In addition, we set $N=2M$, which implies  $\theta_{j+1}-\theta_j=\phi_{i+1}-\phi_i$ and leaves isoceles, right triangles at the equator. Then, we compute numerically $\Psi$ by direct summation of the individual contributions $\delta \Psi_{ijk}$, according to Eq. \eqref{eq:psitotal}. This is the ``direct method'' in the following. We could compare potential values to reference ones in several meridional planes, which is relatively tedious. Instead, the comparison is made along a few test-lines defined as
\begin{equation}
  {\mathbf r}(k)=\frac{2k}{K}\mathbf{OV},
  \label{eq:kspace}
  \end{equation}
where $K$ and $k \in [0,K]$ are integers, point O is the center of coordinates, and V$({\mathbf r})$ is a vertex in the sample. With this prescription, the field point P$({\mathbf r})$ is
\begin{itemize}
  \setlength{\itemsep}{+0pt}
\item inside the body when $2k<K$,
\item just at a vertex for $2k=K$,
\item outside the body when $2k>K$.
\end{itemize}
For these three solids, the potential $\psiref$ is exactly known, and it can therefore serve as a reference \citep[see, e.g.,][]{binneytremaine87}. In order to compare the potential with this reference, we define the error index
\begin{equation}
  E=\log \left|1-\frac{\Psi}{\psiref}\right|,
  \label{eq:errorindex}
\end{equation}
and $-E$ is basically the number of correct digits. Because the sphere, the spheroid and the triaxial have all two planes of symmetry, we do not visit all the vertices with Eq. \eqref{eq:kspace} to test the method. First, we take $j \in \{0,1,\dots,\frac{M}{2}\}$, which corresponds to the upper-plane $z \ge 0$ ($j=0$ is for the pole while $j=M/2$ is for the equator). Second, we limit $\phi$ to values in the range $[0,\pi]$, which is performed by setting $i=j$. This represents $\frac{M}{2}+1$ test-lines in total. So, for a discretization level $\ell=5$, we have $N=32$ and $M=16$, leaving $T=960$ triangles and $9$ test-lines, which are visible in Fig.~\ref{fig:sphere_testlines.pdf} for the sphere. We show in Fig.~\ref{fig:test_sphere.pdf} the error index $E$ computed along these test-lines for the three geometries, where we have set $a=1$, $c=a/2$ and $b=3a/4$. The radial sampling is performed with $k \le K$ and $K=32$ in Eq. \eqref{eq:kspace}. We observe that, whatever the geometry, the error is slightly smaller inside that body than outside it, but it remains relatively uniform, of the order of $0.03$ in relative. This is not really satisfactory yet, but that resolution is very  low. We have also noticed that the results are unchanged if, in Eq. \eqref{eq:kspace}, point V is replaced by the center C of the triangle and for the middle of edges. This confirms that the triangle's edges, its center and its vertices represent fully regular field points.

\begin{figure}[h]
       \centering
       \includegraphics[height=7.5cm,trim={6.5cm 0.cm 6.1cm 1.1cm},clip]{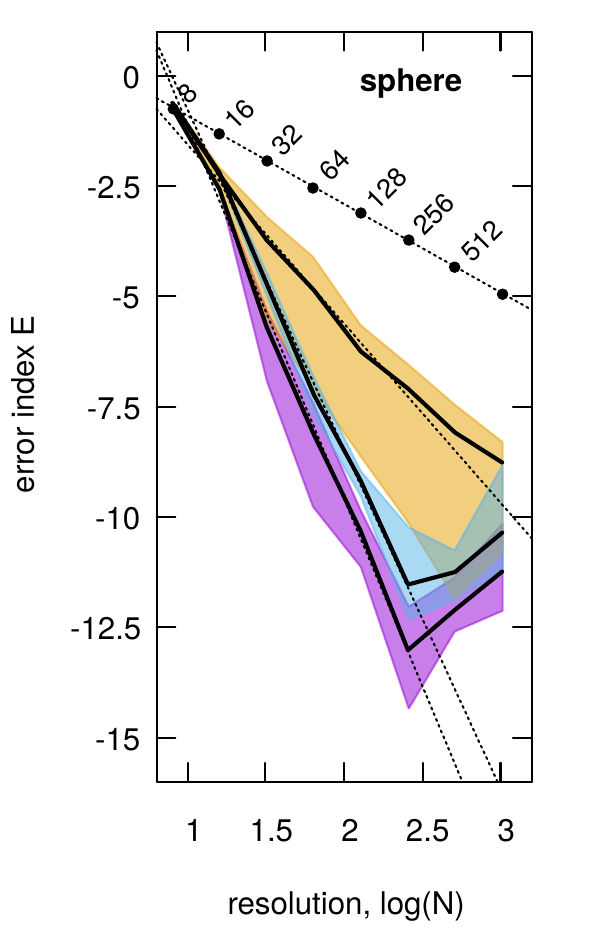}\includegraphics[height=7.5cm,trim={9.1cm 0.cm 6.1cm 1.1cm},clip]{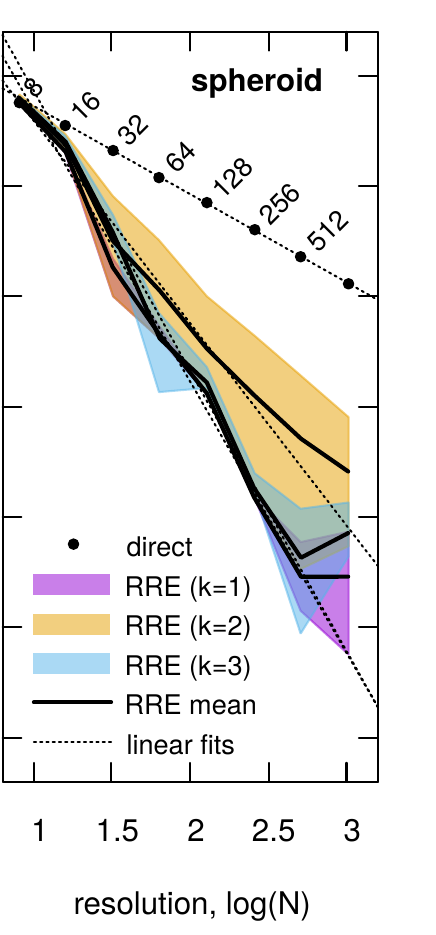}\includegraphics[height=7.5cm,trim={9.1cm 0.cm 7.cm 1.1cm},clip]{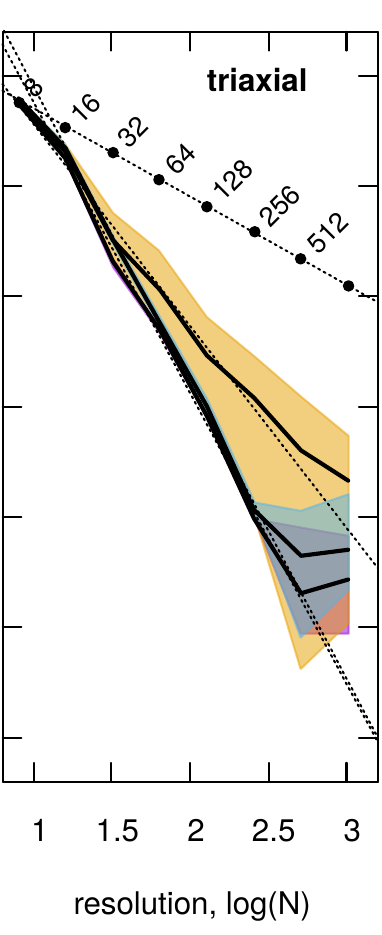}
       \caption{Error index $E$ computed by direct summation ({\it black dots}), i.e., from Eq. \eqref{eq:psitotal}, for the trianguled sphere  with unit radius ({\it left}). Labels $8$, $16$,..., $512$ refer to $N=2M$.  With direct summation, the convergence is second-order; see Table~\ref{tab:slopes} for the fits. When applying a Repeated Richardson Extrapolation ({\it bold lines and colored zones}), which uses the evaluation of the potential on coarser grids, the precision obtained at the same resolution is boosted (see Sect.~\ref{sec:rre}). We have considered three regions: inside the sphere ($k=1$; {\it purple}), at the vertices ($k=2$; {\it brown}) and outside the sphere ($k=3$; {\it turquoise}), according to Eq. \eqref{eq:kspace}  fed with $K=4$. The results obtained for the spheroid and the triaxial ellipsoid are also shown ({\it middle and right})}
       \label{fig:test_RRE_sphere.pdf}
\end{figure}

\subsection{Effect of the numerical resolution}

We have repeated the tests presented above by varying the numerical resolution. Changing $\ell$ modifies the number of triangles drastically, decreases their size and pushes all the vertices closer to the circumscribed surface of the body. As we increase the resolution and change $\ell \rightarrow \ell+1$, each triangle is replaced by four smaller triangles. For these new tests, we use $\ell=3$ for the coarsest grid and $\ell=10$ for the finest one, which corresponds to $N=1024$ and $T=1\,046\,528$ triangles at most. We consider the same test-lines but with the parameters $K=4$ and $k \in \{1,2,3\}$ in Eq. \eqref{eq:kspace}, which limits the radial sampling to only three points, namely one point inside, one point at a vertex (i.e., the surface), and one point outside the body. We show in Fig.~\ref{fig:test_RRE_sphere.pdf} the variation of the error with $N$ for the sphere, the spheroid and the triaxial. We first see that we get a $5$-digit accuracy with $\ell=10$ inside the body, at a vertex and outside the body. The error index is very well represented by a linear fit 
\begin{equation}
  E(N) = \alpha \log N + \beta,
  \label{eq:eindex}
\end{equation}
where $\alpha$ is reported in Table~\ref{tab:slopes} (rows 1 to 3). Clearly, the convergence rate is second-order with respect to $\log N$. It is therefore second order with respect to the number of triangles, and with respect to the size of triangles.

 \begin{table}
   \centering
   \caption{Coefficient $\alpha$ of the linear fit of the error index $E(N)$ via Eq.\eqref{eq:eindex} for the six test-geometries when using the direct method}
   \begin{tabular}{llllll}
                   & Inside ($k=1$)   & Surface ($k=2$) & Outside ($k=3$) &\\\hline
            Sphere & $-1.999 \pm 0.000$  & $-1.996 \pm 0.001$ & $-1.998 \pm 0.001$\\
          Spheroid & $-1.996 \pm 0.001$  & $-1.994 \pm 0.001$ & $-1.995 \pm 0.002$\\
Triaxial ellipsoid & $-1.997 \pm 0.001$  & $-1.995 \pm 0.002$ & $-1.996 \pm 0.002$\\
           Dumbell & $-1.997 \pm 0.001$  & $-1.994 \pm 0.002$ & $-1.996 \pm 0.001$\\
       Lemon-shape & $-2.000 \pm 0.000$  & $-1.997 \pm 0.001$ & $-2.000 \pm 0.000$\\
             Torus & $-1.998 \pm 0.001$  & $-1.996 \pm 0.001$ & $-1.997 \pm 0.001$\\ \\
          433 Eros & $-2.094 \pm 0.055$  & $-2.092 \pm 0.055$ & $-2.071 \pm 0.054$ \\ \hline
    \end{tabular}
   \label{tab:slopes}
    The statistics are performed by omitting the first two points, i.e. for $N \ge 32$. The last row is for 433 Eros (see Sect.~\ref{sec:eros}).
 \end{table}
 
\begin{table}[h]
  \caption{Spherical radius $r(\theta,\phi)$ for the three geometries shown Fig.~\ref{fig:testnewshapes.pdf}, and specific values of the coefficients $a$, $b$ and $c$}
  \centering
  \begin{tabular}{lll}\hline \hline
  Shape    & Spherical radius $r$ & Comment \\  \hline
  Dumbell  & $r=a+b\cos(2\theta)$ & $a=1$, $b=\frac{1}{2}$\\
  Lemon-shape    & $r \sqrt{(1+a \sin \theta)^2+\cos^2 \theta} =c$ & $a=2$, $c=\sqrt{2}$\\
  Torus    & $r^2=a^2+b^2+2ab\cos 2\theta$ & $a=\frac{3}{4}$, $b=\frac{1}{4}$ \\ \hline
  \end{tabular}
  \label{tab:rnewshapes}
  In all cases, $\theta \in [0,\pi]$ and $\phi \in [0,2\pi]$. Regarding the torus, $\theta$ is defined by $\sin \theta = \mathbf{\Omega V} \cdot \mathbf{e}_z$ where $\Omega(a \cos \phi,a \sin \phi,0)$ is the centre of the meridional section.
\end{table}

\begin{figure}[h]
       \centering
       \includegraphics[height=6.cm,trim={7.5cm 3.5cm 10cm 3.5cm},clip]{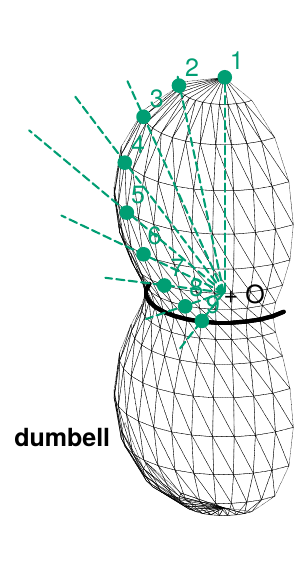}\qquad\includegraphics[height=6.cm,trim={8.5cm 3.5cm 9.3cm 3.5cm},clip]{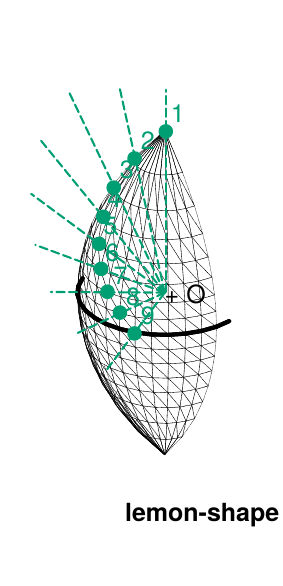}\qquad\includegraphics[height=6.cm,trim={8cm 3.5cm 7.5cm 3.5cm},clip]{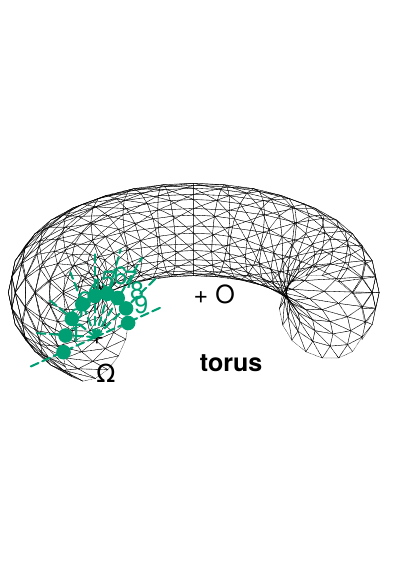}\\
       \caption{Surface triangulation of the dumbell ({\it left}), the lemon-shape surface  ({\it middle}) and the torus ({\it right}) obtained with $N=2M=32$, and the $9$ test-lines used to test the method}
       \label{fig:testnewshapes.pdf}
\end{figure}

\begin{figure}[h]
       \centering
       \includegraphics[height=7.5cm,trim={6.8cm 0.cm 6.3cm 1.1cm},clip]{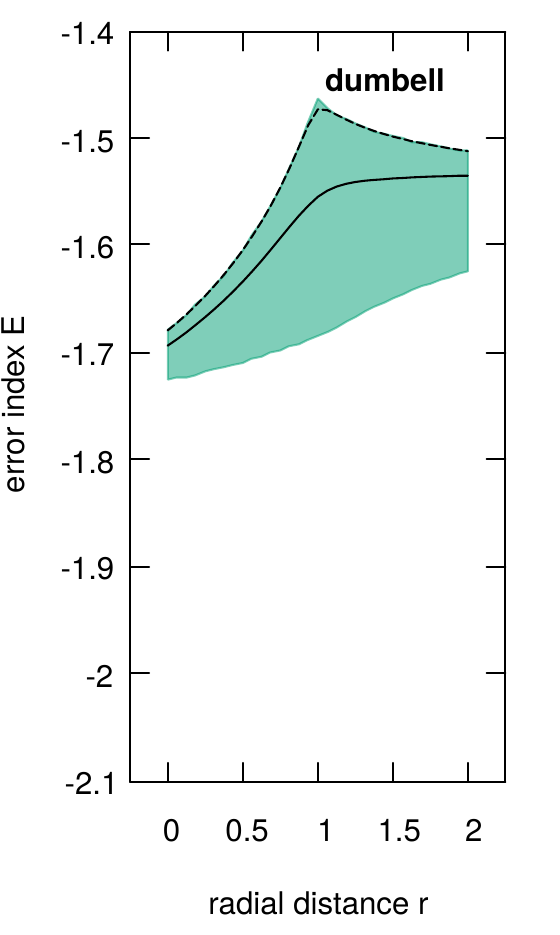}\includegraphics[height=7.5cm,trim={9cm 0.cm 6.3cm 1.1cm},clip]{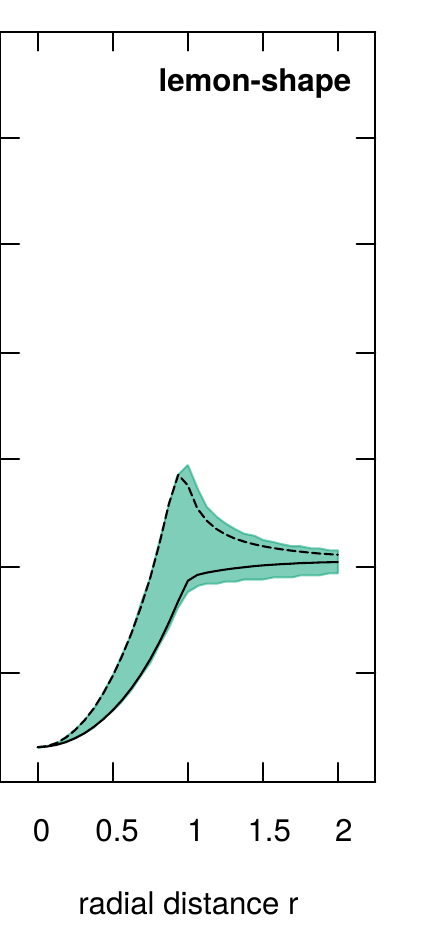}\includegraphics[height=7.5cm,trim={9cm 0.cm 7.2cm 1.1cm},clip]{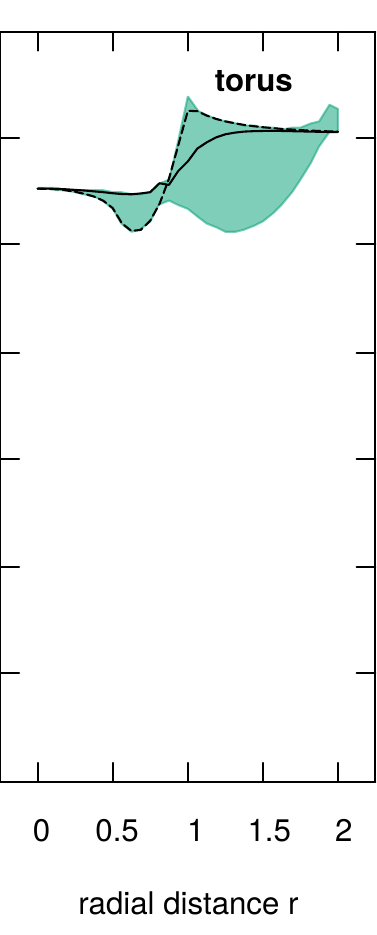}
       \caption{Same caption and same conditions as for Fig.~\ref{fig:test_sphere.pdf} but for the dumbell ({\it left}), for the lemon-shape surface ({\it middle}) and for the torus ({\it right}); see Fig.~\ref{fig:sphere_testlines.pdf} for the sphere}
       \label{fig:test_others.pdf}
\end{figure}

\subsection{Three test-geometries with unknown reference}

As new test-bodies, we have considered three new geometries, namely
\begin{itemize}
\setlength{\itemsep}{+0pt}
\item a dumbell,
\item a lemon-shape surface,
\item a torus with circular sections,
\end{itemize}
again with triangulated surface based on spherical coordinates (see above). The equations $r(\theta,\phi)$ are given in Table~\ref{tab:rnewshapes} and the shapes are displayed in Fig.~\ref{fig:testnewshapes.pdf}. Note that last two geometries are particularly interesting: the poles of lemon-shape are sharp, and the torus is not a simply connected body. The potential being unfortunately not known for these new shapes, the reference is defined as the potential computed at a higher resolution. As the convergence rate is second-order, the reference obtained with $8 N$ is sufficient, and so we take $\psiref \equiv \Psi(8N)$.

\begin{figure}[h]
       \centering
       \includegraphics[height=7.5cm,trim={6.5cm 0.cm 6.1cm 1.1cm},clip]{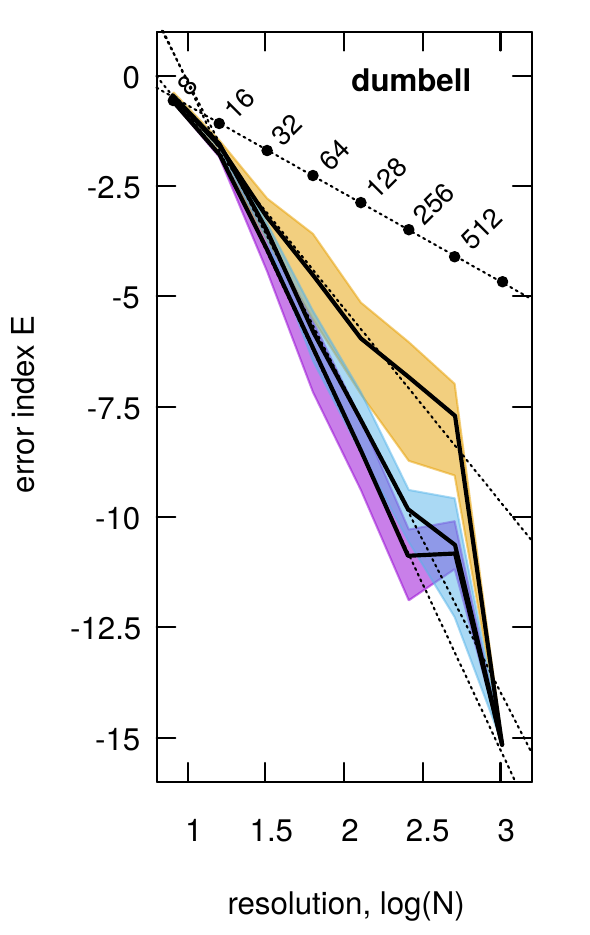}\includegraphics[height=7.5cm,trim={9.1cm 0.cm 6.1cm 1.1cm},clip]{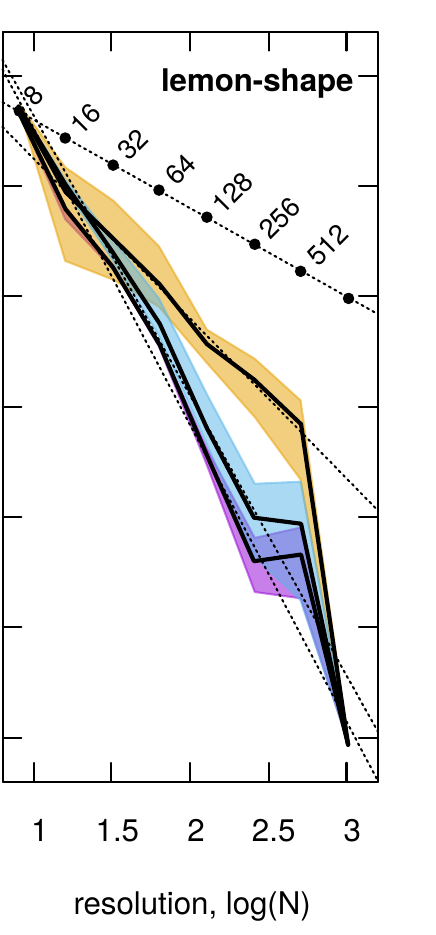}\includegraphics[height=7.5cm,trim={9.1cm 0.cm 7.cm 1.1cm},clip]{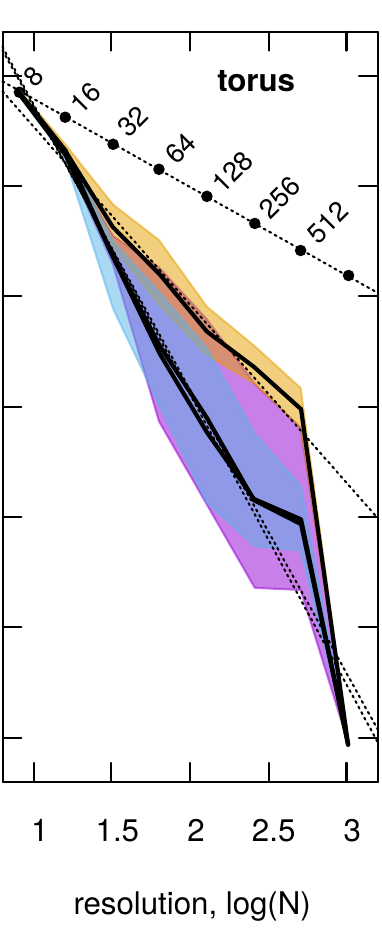}
       \caption{Same caption and same conditions as for Fig.~\ref{fig:test_RRE_sphere.pdf} but for the dumbell ({\it left}), for the lemon-shape surface ({\it middle}) and for the torus ({\it right})}
       \label{fig:test_RRE_dumbell.pdf}
\end{figure}

We have performed the same series of tests as above, first at low resolution along $9$ test-lines, and second at three points by varying the spatial resolution. For the torus, the test-lines run from the center $\Omega$ of the local meridional section (see the figure), i.e., we use Eq. \eqref{eq:kspace} but $\mathbf{OV}$ is replaced by $\mathbf{\Omega V}$. Figure~\ref{fig:test_others.pdf} shows the error index $E$ along the 9 test-lines for the dumbell, the lemon-shape surface and the torus. The evolution of this index with the numerical resolution $N$ for $k \in \{1,2,3\}$ and $K=4$ is displayed in Fig.~\ref{fig:test_RRE_dumbell.pdf}. We give in Table~\ref{tab:slopes} (rows 4 to 6) the coefficient $\alpha$ obtained by linearly fitting $E$ versus $\log N$ via Eq. \eqref{eq:eindex}. All the results are very similar as those obtained for the sphere, the spheroid and the triaxial. In particular, the convergence is second-order in the mesh spacing, which is ideal for convergence acceleration (see below).

\section{Convergence acceleration by using the Repeated Richardson Extrapolation (RRE)}
\label{sec:rre}

Richardson's Extrapolation (RE) is a standard numerical technique for improving quadratures and derivatives. It uses two evaluations of a given quantity obtained at two different resolutions. A subtle combination of the same quantity obtained at these resolutions greatly improves the accuracy, provided the quantity is sufficiently smooth or regular. The process is repeated for a series of increasing resolutions to achieve a desired accuracy. The ratio of sucessives spacings is typically $2$. This is the Repeated Richardson Extrapolation (RRE); see the Appendix \ref{app:rre}.

 As the determination of the potential values from surface triangulation is clearly a $2$nd-order quadrature in the mesh spacing, applying the RRE technique is fully appropriate in the present situation. In practice, for a given geometry or body, we first generate a series of nested grids starting with $N=2^3=8$ (the coarsest grid), then $N=2^4=16$, ... up to $N=2^{10}=1024$ (the finest grid), which therefore correspond to $\ell \in [3,10]$; see Sect.~\ref{subsec:test3geometries}. Next, we select a point P$(\mathbf{r})$ in space and we compute the potential $\Psi(\mathbf{r})$ from Eq. \eqref{eq:psitotal} for each grid. Values are slightly different to each other, and are denoted $\Psi_3(\mathbf{r})$, $\Psi_4(\mathbf{r})$, $\dots \Psi_{10}(\mathbf{r})$, and we set
\begin{equation}
  A_{\ell,3}(\mathbf{r}) = \Psi_\ell(\mathbf{r}), \qquad \text{with } \ell \in [3,10].
\end{equation}
Next, we compute all the elements $A_{\ell,m}(\mathbf{r})$ for $m \in [4,\ell]$ according to Eq.\eqref{eq:richardsonextrapolationtable} of Appendix \ref{app:rre}. The element obtained for $m=\ell$, i.e., $A_{\ell,\ell}$, is the improved estimate for $\Psi(\mathbf{r})$. This technique works for any field point P$(\mathbf{r})$.

\begin{table}[h]
  \caption{Potential at the center of the homogeneous sphere with unit radius (column 3) computed with the RRE technique in double precision, to be compared to the exact value $\psiref(\mathbf{0}) = - 2 \pi$}
  \begin{tabular}{llllll}
$\ell$ &  $A_{\ell,3}$            & $A_{\ell,4}$             & $A_{\ell,5}$            & $A_{\ell,6}$       \\ \hline
 3&  -5.2658724606271727 \\
 4&  $\underline{-6}.0177824547115533$ & $-6.2684191194063468$ \\
 5&  $\underline{-6.2}161323588550710$ & $\underline{-6.28}22489935695769$ & $\underline{-6.2831}709851804591$ \\
 6&  $\underline{-6.2}663780221959700$ & $\underline{-6.2831}265766429363$ & $\underline{-6.2831850}821811601$ & $\underline{-6.283185305}9430758$ \\
 7&  $\underline{-6.2}789807304638492$ & $\underline{-6.28318}16332198089$ & $\underline{-6.28318530}36582670$ & $\underline{-6.28318530717}37767$  \\
 8&  $\underline{-6.28}21339907476057$ & $\underline{-6.283185}0775088576$ & $\underline{-6.283185307}1281270$ & $\underline{-6.2831853071}832038$  \\
 9&  $\underline{-6.28}29224673155020$ & $\underline{-6.283185}2928381338$ & $\underline{-6.283185307}1934187$ & $\underline{-6.2831853071}944552$  \\
 10&  $\underline{-6.2831}195965837843$ & $\underline{-6.28318530}63398784$ & $\underline{-6.2831853072}399948$ & $\underline{-6.283185307}2407338$  \\ \\ 
$\ell$   & $A_{\ell,7}$             & $A_{\ell,8}$            & $A_{\ell,9}$             & $A_{\ell,10}$\\ \hline
 7&  $\underline{-6.28318530717}86030$ \\
 8&  $\underline{-6.2831853071}832402$ & $\underline{-6.2831853071}832446$ \\
 9&  $\underline{-6.2831853071}944996$ & $\underline{-6.2831853071}945112$ & $\underline{-6.2831853071}945138$ \\
 10&  $\underline{-6.283185307}2409158$ & $\underline{-6.283185307}2409611$ & $\underline{-6.283185307}2409727$ & $\underline{-6.283185307}2409753$  \\ \\
 &&& reference (exact)  &   $-6.2831853071795864$ \\\hline
 \end{tabular}
  \label{tab:RREatthecenter}
  The table of elements $A_{\ell,m}$ is constructed according to Eq. \eqref{eq:richardsonextrapolationtable}. The second column contains $A_{\ell,3}$, which corresponds to $\Psi_3(\mathbf{0})$, $\Psi_4(\mathbf{0})$, $\dots \Psi_{10}(\mathbf{0})$. See the appendix,  Table~\ref{tab:RREatthecenter_qp}, for quadruple precision.
\end{table}

\subsection{Example at the centre of the sphere}

 Table~\ref{tab:RREatthecenter} lists potential values obtained at the center of the sphere with unit radius (the exact value is $-2\pi$) for all discretization levels $\ell \in [3,10]$, as well as all the elements $A_{l,k}$ of the table. This field point being inside the body, the RRE is especially efficient in drastically reducing the error. We observe that, from $\ell=8$, the error increases with $\ell$, which is a well-known effect of arithmetic with computers. When the amount of surface triangles become extremely large, substration cancellations and round off errors inevitably inhibit onvergence \citep[e.g.,][]{numrec92,f17}. We have verified that the this effect disappears with quadruple precision; see Tab.~\ref{tab:RREatthecenter_qp} in the Appendix.

 \begin{table}
   \centering
   \caption{Same caption as for Table~\ref{tab:slopes}, but after applying the Repeated Richardson Extrapolation (RRE) in columns 2, 4 and 6; see also Figs.~\ref{fig:test_RRE_sphere.pdf} and \ref{fig:test_RRE_dumbell.pdf}}
   \begin{tabular}{llllllll}
            & $\alpha_{\rm RRE}$ (inside)   & $\Delta \alpha$      & $\alpha_{\rm RRE}$ (surface) &  $\Delta \alpha$ & $\alpha_{\rm RRE}$ (outside) & $\Delta \alpha$\\\hline
Sphere      & $-8.461 \pm 0.224$ & $\mathbf{6.5}$ & $-4.313 \pm 0.198$ & $\mathbf{2.1}$ & $-7.403 \pm 0.204$ & $\mathbf{5.7}$\\
Spheroid    & $-6.147 \pm 0.423$ & $\mathbf{4.1}$ & $-4.582 \pm 0.386$ & $\mathbf{2.6}$ & $-6.344 \pm 0.405$ & $\mathbf{4.3}$\\
Triaxial    & $-6.584 \pm 0.241$ & $\mathbf{4.6}$ & $-4.569 \pm 0.321$ & $\mathbf{2.6}$ & $-6.700 \pm 0.151$ & $\mathbf{4.7}$\\
dumbell     & $-7.552 \pm 0.085$ & $\mathbf{5.6}$ & $-4.385 \pm 0.246$ & $\mathbf{2.4}$ & $-6.920 \pm 0.112$ & $\mathbf{4.9}$\\
Lemon-shape & $-6.713 \pm 0.490$ & $\mathbf{4.7}$ & $-3.620 \pm 0.139$ & $\mathbf{1.6}$ & $-6.339 \pm 0.319$ & $\mathbf{4.3}$\\
Torus       & $-6.413 \pm 0.231$ & $\mathbf{4.4}$ & $-4.031 \pm 0.285$ & $\mathbf{2.0}$ & $-6.566 \pm 0.359$ & $\mathbf{4.6}$\\\\
433 Eros    & $-2.541 \pm 0.249$ & $\mathbf{0.4}$ & $-2.485 \pm 0.245$ & $\mathbf{0.4}$ & $-2.556 \pm 0.250$ & $\mathbf{0.5}$ & see Sect.~\ref{sec:eros}\\ \hline 
    \end{tabular}
  \label{tab:slopes_RRE}
  The statistics are performed by omitting the first point and points beyond the knee, i.e. for $32 \le N \le 256$. The quantity $\Delta \alpha$ in columns 3, 5 and 7 ({\it bold}) represent the improvement in accuracy for every ten-fold increase in $N$ with respect to the direct method, i.e. $\alpha - \alpha_{\rm RRE} \equiv \Delta \alpha$.
 \end{table}
 
\subsection{Revisiting the six test-geometries}

We show in Figs.~\ref{fig:test_RRE_sphere.pdf} and \ref{fig:test_RRE_dumbell.pdf} the potential computed along test-lines when using the Repeated Richardson Extrapolation, compared with the direct method. We see that the gain in the precision is large outside and inside the body. In particular, the precision of the direct method obtained for $N=1024$ is reached with only $N=64$ with the RRE technique. The computing time is therefore decreased by about $(1024/64)^4$, theoretically (see Sect.~\ref{sec:discussion}). Table~\ref{tab:slopes_RRE} (rows 1 to 6) gives the coefficients $\alpha_{\rm RRE}$ obtained by fitting the error index $E_{\text{RRE}}$, according to Eq. \eqref{eq:eindex}. We find $\alpha_{\rm RRE} \lesssim -6$ on average for six geometries (inside and outside matter), which means a gain $\Delta\alpha \equiv \alpha - \alpha_{\rm RRE} \approx 4$. At the surface, the results are less remarkable but still very satisfactory, with $\alpha_{\rm RRE} \lesssim -4$ and $\Delta \alpha \approx 2$ (i.e. twice  extra digits fixed with the RRE method).

\begin{figure}[h]
       \centering
       \includegraphics[height=5.cm,trim={2.5cm 4cm 0cm 3.5cm},clip]{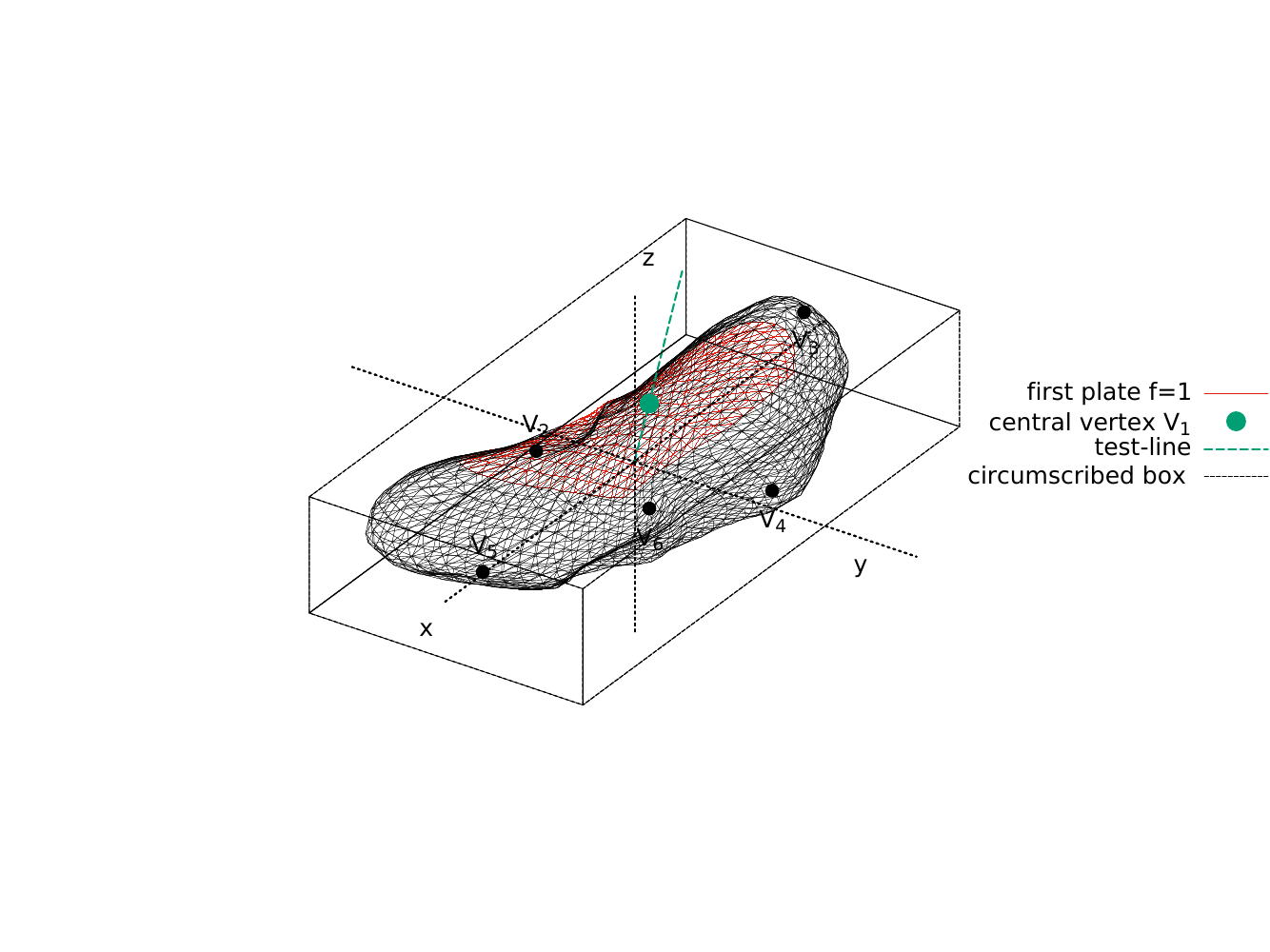}
       \caption{Surface of asteroid 433 Eros by using data from the NASA Data Planetary System \citep{gaskell21}. The resolution corresponds to $q=16$, leaving $3072$ triangles in total. The first plate $f=1$ is colored ({\it red}) and the test-line passing through the central vertex V$_1$ are marked ({\it green})}
       \label{fig:eros_scheme.pdf}
\end{figure}

 \subsection{Real case. Example with asteroid 433 Eros}
 \label{sec:eros}

 As shown in the previous sections, the performance of the RRE method are quite similar for all geometries, with an advantage for the sphere. For this reason, we decided to continue exploring the performance of the method with a relatively irregular, really non-spherical object. Undoubtedly, real asteroids have no smooth surface, and disapointed results are expected. Figure~\ref{fig:eros_scheme.pdf} shows the surface of 433 Eros reproduced from the ASCII-files available at NASA Planetary Data System \citep{gaskell08,gaskell21}. The topographic data, deduced from NEAR MSI images, are stored according to the implicitly connected quadrilateral (ICQ) format, which consists of $6$ square plates, labeled with $f \in [1,6]$, and containing $q+1$ nodes per direction. Each node has ICQ-coordinates V$_{ijf}(i,j,f)$ and Cartesian coordinates V$_{ijf}(x_{ijf},y_{ijf},z_{ijf})$. The unit is kilometer, and there are $7$ significant figures (see below). The finest grid available is for $q=512 \equiv Q$, leaving $T=12 \times 512^2$ triangles in total, and this is the grid we use in the following. This format is especially appropriate regarding the present purpose as each quadruplet of adjacent nodes is already arranged into $2 q^2$ triangles per plate. More importantly, one gets easily nested grids by skipping $1$ node (with $q=Q/2$), $3$ nodes (with $q=Q/4$), $7$ nodes (with $q=Q/8$) etc. per direction.
 In practice, we divide all Cartesian coordinates by a length scale $L$. We define $L = \max \{ L_x,L_y,L_z\}$, where
\begin{flalign}
  L_x = \max\{|x_{ijk}|\}_{{\rm all \, triangles \,} T},
\end{flalign}
and similar definitions for $L_y$ and $L_z$. Next, we assume a constant mass-density $\rho$ for Eros. Potential values are therefore expressed in units of $G \rho L^2$. From the data, the circumscribed box is $33$ km $\times \, 17$ km $\times \, 12$ km and we find $L = L_x = 17.57339$ km, $L_y/L \approx 0.490$ and $L_z/L \approx 0.346$, and a mean spherical radius of about $0.5467 \, L$. The adimensionned surface area and mass are found to be ${\cal A}/L^2 \approx 3.6864$ and $M/\rho L^3 \approx  0.4620$, respectively. The centre of coordinates is almost the center of mass G, as we find G$(-1.47\times 10^{-5},-1.26\times 10^{-5}, +6.48\times 10^{-5}) $ in units of $L$. The typical geometrical parameters of the triangles are listed in Table~\ref{tab:erostijdata} for various grid parameter $q$.

 \begin{table}[h]
   \centering
   \caption{Typical geometrical parameters of the triangles versus the resolution $q$ for 433 Eros : mean triangle surface area (column 3) and mean length of triangles (columns 4 and 5). Data are from \cite{gaskell21}}
   \begin{tabular}{rrrrrl}
     $q$ & $T$       & ${\cal A}/L^2 T$       & $\sqrt{2\langle {\cal A} \rangle_T}$  (m) & $\langle t_{ij}\rangle_T$ (m) & Comment \\  \hline
   $512$ & $3145728$ & $1.2172 \times 10^{-6}$ & $26.9$ & $32.8$ & $q=Q$ \\
   $256$ & $786432$  & $4.6840 \times 10^{-6}$ & $53.8$ & $65.6$ \\
   $128$ & $196608$  & $1.8695 \times 10^{-5}$ & $107.5$ & $131.0$ \\
    $64$ & $49152$   & $7.4439 \times 10^{-5}$ & $214.4$ & $261.2$ \\
    $32$ & $12288$   & $2.9579 \times 10^{-4}$ & $427.4$ & $520.2$\\
    $16$ & $3072$    & $1.1730 \times 10^{-4}$ & $851.2$ & $1035.6$ & $\triangleleft$ see Fig.~\ref{fig:eros_scheme.pdf}\\\hline
   \end{tabular}
  \label{tab:erostijdata}
 \end{table}
 
\begin{figure}
       \centering
       \includegraphics[width=8.9cm,trim={3cm 0.5cm 2.5cm 3cm},clip]{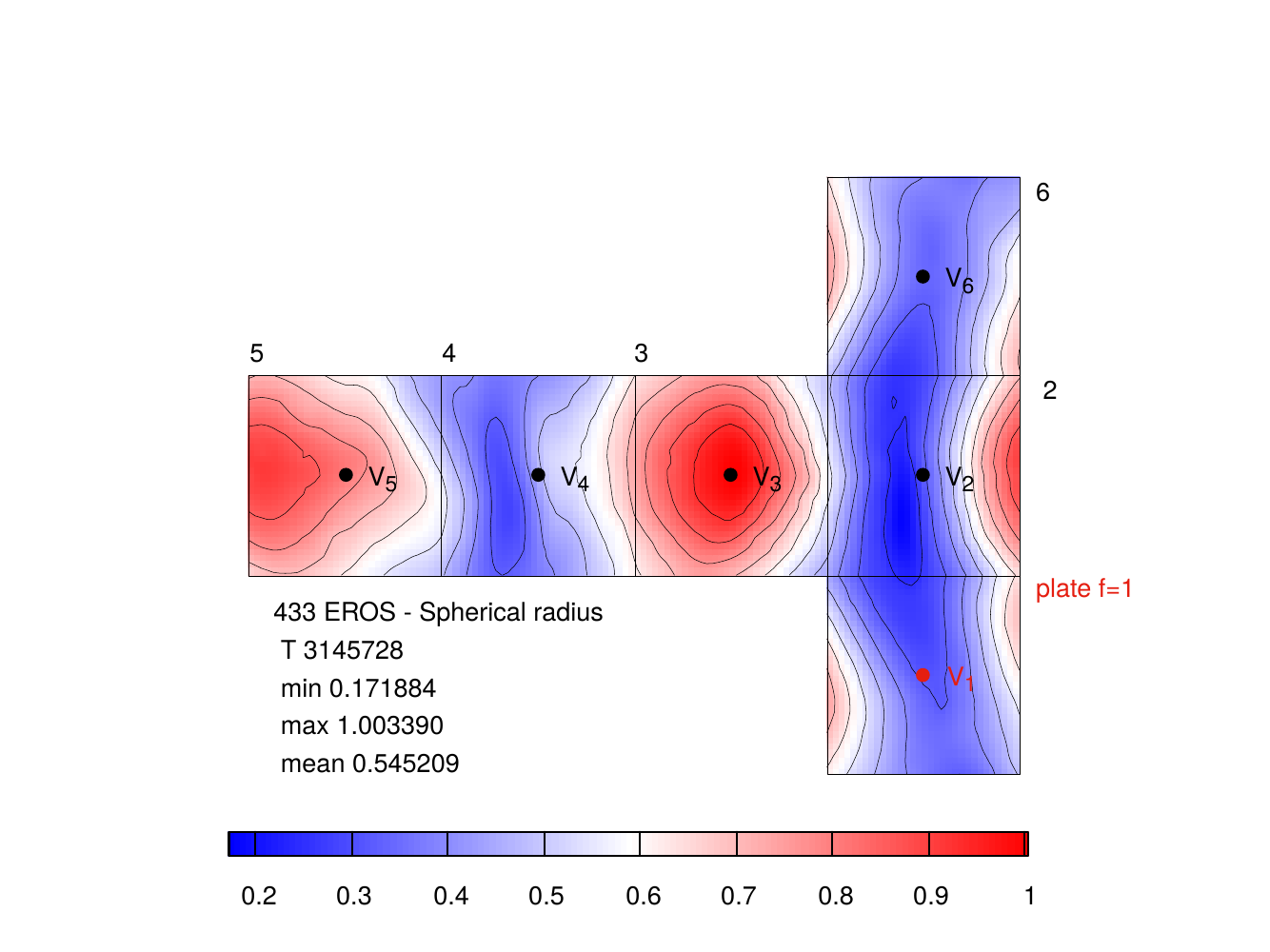}
       \includegraphics[width=8.9cm,trim={3cm 0.5cm 2.5cm 3cm},clip]{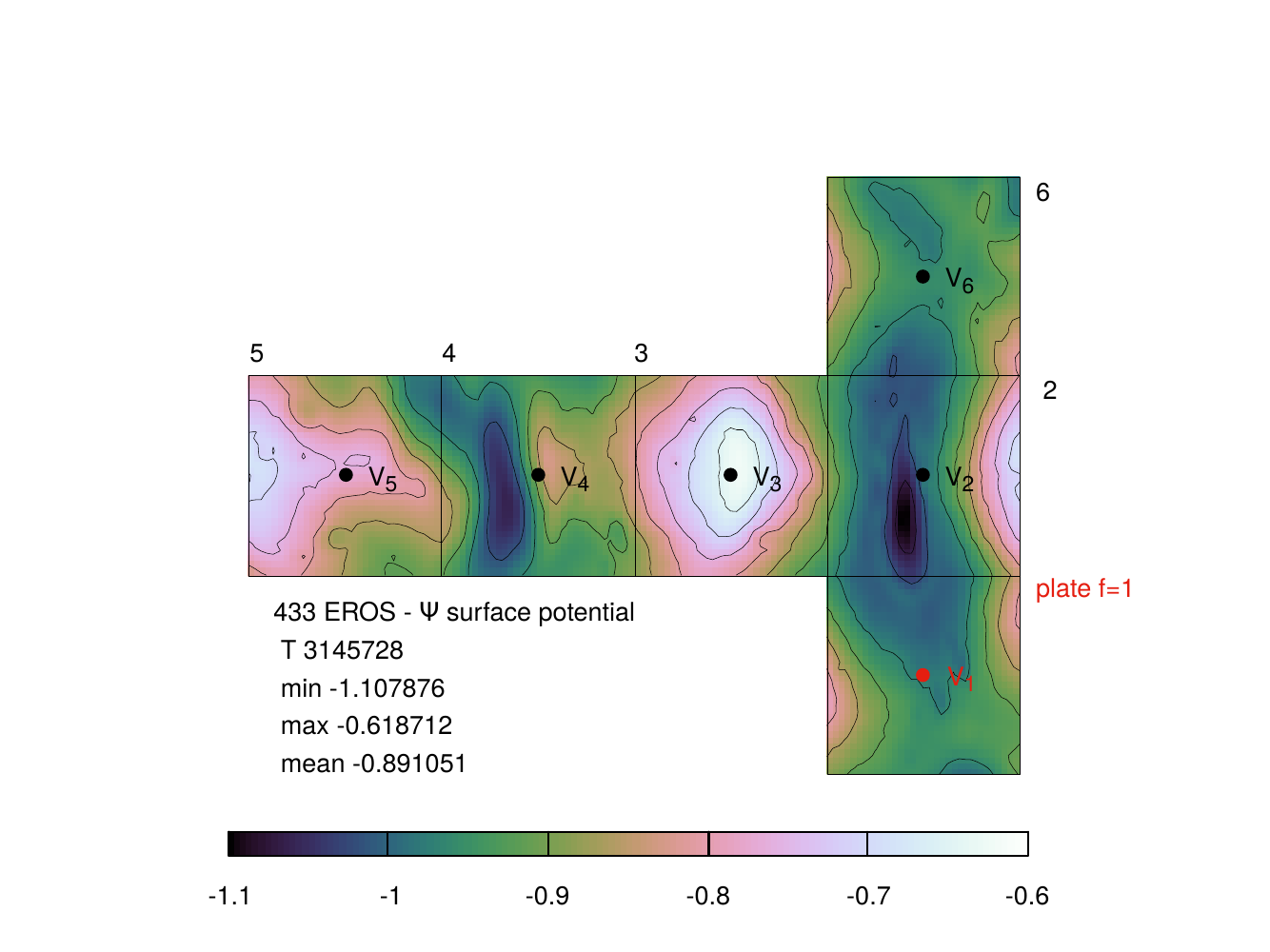}
       \caption{Spherical radius $r/L$ of asteroid 433 Eros ({\it top panel}) and potential $\Psi/G \rho L^2$ at the surface  ({\it bottom panel}) by using data from the NASA Data Planetary System \cite{gaskell21}. The first panel $f=1$ is colored ({\it red}) and the central vertex V$_1$ is marked; see also Fig.~\ref{fig:eros_scheme.pdf}.}
       \label{fig:eros16_16.pdf}
\end{figure}

We show in Fig.~\ref{fig:eros16_16.pdf} the normalized spherical radius $\sqrt{x^2+y^2+z^2}/L$, and the potential $\Psi$ computed at full resolution $q=Q$ by direct summation\footnote{In Sect.~\ref{sec:eros}, we limit the computations to the vertices of a coarse grid corresponding to $q=Q/32=16$. This saves a lot of computing time which goes like $\tau(N) \propto N^4$. This choice is not critical for the present purposes \label{note:coarsegrid}}, which constitutes the reference, $\Psi_Q \equiv \Psi_{\rm ref}$. Mean values for the potential and for the acceleration at the surface are 
\begin{flalign}
  \begin{cases}
    \frac{1}{G \rho L^2}\langle \Psi \rangle_T \approx -0.891\\
    \frac{1}{G \rho L^3}\langle  \left| {\mathbf \nabla} \Psi \right| \rangle_T \approx  1.684.
  \end{cases}
  \label{eq:psigmean}
\end{flalign}

 \begin{table}[h]
   \centering
   \caption{Potential at the test-vertex V$_1$ located at the center of the first plate $f=1$ obtained at different resolutions (column 3) and mean contribution of triangles  $\langle \delta \Psi_{ijk} \rangle$ (column 4)}
   \begin{tabular}{rrrcl}
     $q$ & $T$       &  $\Psi/G\rho L^2$ at V$_1$ & $\langle \delta \Psi_{ijk} \rangle /G\rho L^2$ & Comment \\  \hline
   $512$ & $3145728$ &  $\underline{-0.9777}9130$ & $-3.1 \times 10^{-7}$ & $q=Q$ \\
$256$ & $786432$     &  $\underline{-0.9777}7931$ & $-1.2\times 10^{-6}$ &\\
$128$ & $196608$    & $\underline{-0.9777}3042$ & $-5.0\times 10^{-6}$ & \\
$64$ & $49152$      & $\underline{-0.977}50019$ & $-2.0\times 10^{-5}$ &\\
$32$ & $12288$      &  $\underline{-0.97}681906$ & $-7.9\times 10^{-5}$ &\\
$16$ & $3072$       & $\underline{-0.97}378461$ & $-3.2\times 10^{-4}$  & $\triangleleft$ see Fig.~\ref{fig:eros_scheme.pdf}\\\hline
   \end{tabular}
  \label{tab:psiatf1}
  Values (only $8$ digits are given) are normalized to $G \rho L^2$ where $L = 17.57339$ km (see text). Values of $q$ and $T$ refer to the number of facets of each panel per direction in the ICQ format and total number of triangles, respectively. See also Fig.~\ref{fig:eros_scheme.pdf}
 \end{table}

\begin{figure}[h]
       \centering
       \includegraphics[height=4.8cm,trim={5.5cm 3.5cm 0.cm 4.5cm},clip]{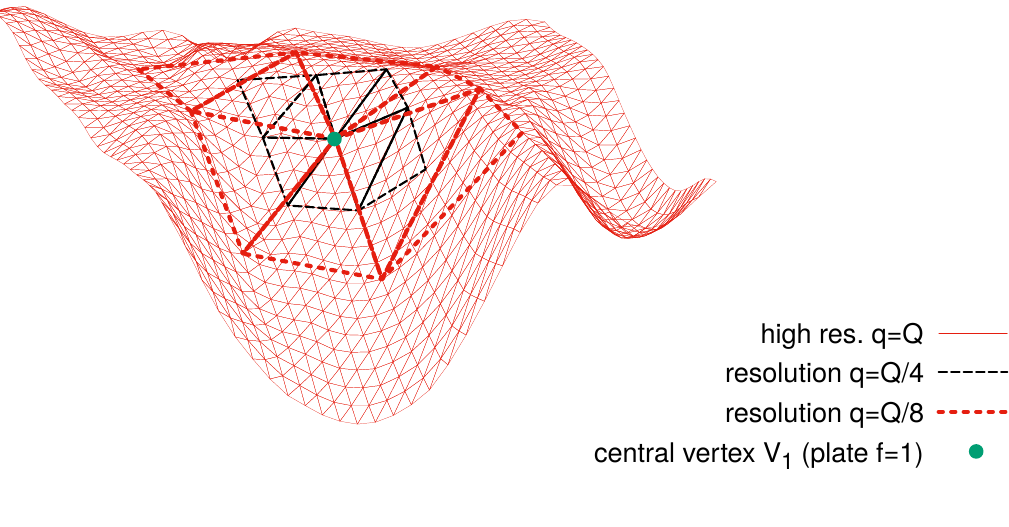}
       \caption{Surface of 433 Eros at three different resolutions zoomed around the central vertex V$_1$ of the first panel $f=1$; see also Fig.~\ref{fig:eros_scheme.pdf}}
       \label{fig:eros_smooth.pdf}
\end{figure}

\begin{figure}[h]
       \centering
       \includegraphics[height=7.5cm,trim={6.5cm 0.cm 6.1cm 1.1cm},clip]{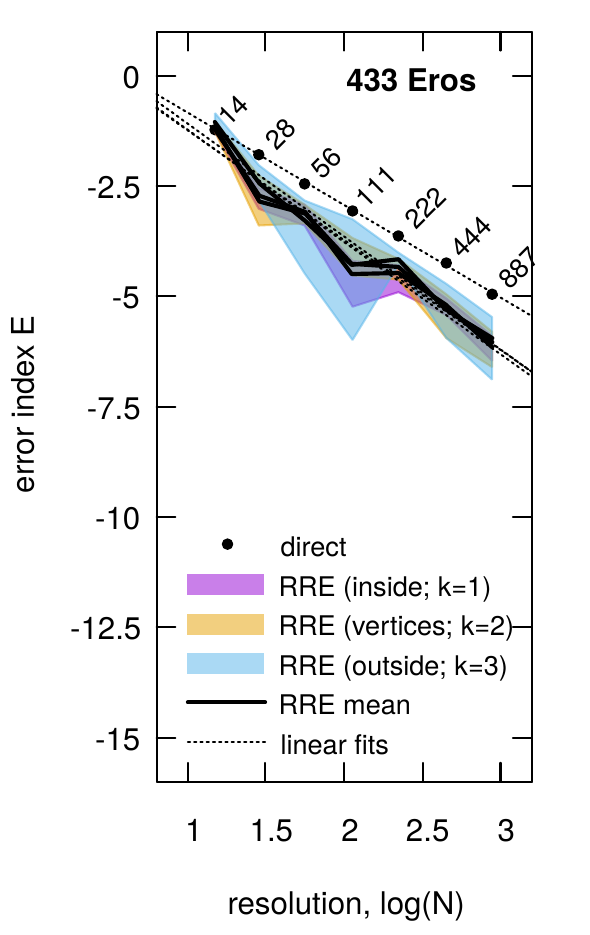}
       \caption{Same caption as for Fig.~\ref{fig:test_RRE_sphere.pdf} but for 433 Eros. Test-lines pass through the central vertices V$_{f=1,6}$, according to Eq. \eqref{eq:kspace} with $K=4$ and $k=\{1,2,3\}$; see also Fig.~\ref{fig:eros_scheme.pdf}.}
       \label{fig:test_RRE_eros.pdf}
\end{figure}

Table~\ref{tab:psiatf1} lists the potential values obtained with the direct method at the central vertex V$_1(\frac{q}{2},\frac{q}{2},1)_{\rm ICQ}$ of the first plate $f=1$ (see Fig.~\ref{fig:eros_scheme.pdf}), at different resolutions $q \in \{Q,Q/2,Q/4, \dots, Q/32\}$, including the finest grid $q=Q$. A zoom of the region around this node is visible in Fig.~\ref{fig:eros_smooth.pdf}. We immediately see from the table that the potential reaches a plateau around $-0.9777$ from $q = 128$, but it does not really converge in the classical sense. The evolution of the number of digits fixed from one resolution to the next one is not really linear in log. scale. As done in Sect.~\ref{sec:rre}, we have considered six test-lines passing through the central vertices V$_f(\frac{q}{2},\frac{q}{2},f)_{\rm ICQ}$, according to Eq. \eqref{eq:kspace} with $K=4$ and $k \in \{1,2,3\}$. This enables to check the method inside and outside the body. Figure~\ref{fig:test_RRE_eros.pdf} shows the error index $E(N)$ obtained by direct summation and with the RRE technique. The slopes are reported in Tables~\ref{tab:slopes} and  \ref{tab:slopes_RRE}, respectively (last row). With direct summation, $\alpha \approx -2$, but this coefficient varies significantly with $N$, as already noticed for the central vertex V$_1$. With $\alpha_{\rm RRE} \approx -2.5$, the benefit of the RRE technique is indeed poor, inside, at the surface and outside the body. This is not a great surprise. Solid bodies considered in the previous sections are ideal, with smooth surfaces and no sub-grid terrain irregularities. The real asteroids's surface lacks smoothness that is required for the RRE method \citep{cheng02,SUSORNEY2018}. Thus, its benefit is modest.

Although the position of the ICQF-nodes is given with $7$ figures for 433 Eros, the relative separation of vertices has significant imprecisions. Errors in the altimetric data are estimated to be of the order of $6$ m for this asteroid, mainly associated with uncertainties of navigation \citep{p24}. \cite{bpm20} have investigated in great details the impact of shape uncertainty on key-quantities, in particular on the gravitational acceleration outside bodies, with an application to 25143 Itokawa whose shape errors are about $20$ m, comparable to Eros, in absolute. This addresses the question of the link between potential values and altimetric errors at the surface (see below).

\section{Impact of altimetric uncertainties}

As shown, the RRE method is very powerful for regular surfaces. Not only asteroids have non-smooth terrains but their shape are also uncertain. The arbitrary displacement of one and only one vertex V$_{ijk} \rightarrow$ V$^{\rm new}_{ijk}$ of the triangulated surface corresponds to generating a bump if V$^{\rm new}_{ijk}$ is outside the initial surface (a cavity otherwise). This operation changes the volume and the mass in a relatively simple manner. It has, however, a less trivial effect on the potential, especially locally, because the new position V$^{\rm new}_{ijk}$ also impacts all relative separations involved in Eq. \eqref{eq:tripleint}. In other words, not only a cavity/bump is created but the location where the potential is evaluated is changed. This is particularly important when studying mechanical processes at the surface of asteroids and below \citep{rkbepg14,bebddhm25}.

\subsection{The prescription}
\label{sec:prescription}

We can estimate the deviation $\Delta \Psi = \Psi({\rm V}^{\rm new}_{ijk}) - \Psi({\rm V}_{ijk})$ from a Taylor expansion, namely
\begin{flalign}
  \Psi({\rm V}^{\rm new}_{ijk}) \approx  \Psi({\rm V}_{ijk})\pm  \lambda \nabla \Psi.
  \label{eq:devpsi}
\end{flalign}
where $\nabla \Psi \equiv g$ is the gravitational acceleration along the direction of the displacement, and $\lambda$ is the height of the bump that can be associated to a shape error ($\lambda\approx 6$ m for 433 Eros). This formula means that any uncertainty in the position of a vertex sets a limit in the precision of the local potential (in units of $G \rho L^2$) of the order of $\lambda/L$. Determining $\Delta \Psi$ from the formula given in Sect.~\ref{sec:solution} is  cumbersome, even for very simple geometries. In fact, $\Delta \Psi$ is roughly the self-gravitating potential of the bump, $\delta \Psi_{\rm self}$, evaluated at the vertex V$^{\rm new}_{ijk}$. If we assume that a vertex V$_{ijk}$ is shared by $4$ similar, right triangles with edges $a_1$ and $a_2$, then moving this vertex V$_{ijk} \rightarrow$ V$^{\rm new}_{ijk}$ by $\lambda >0$ perpendicularly to the local plane forms a pyramidal bump with height $\lambda$, diamond-shape base, and volume $V= \frac{4}{3}a_1a_2 \lambda$. It would be possible to determine $\delta \Psi_{\rm self}$ exactly from Eq. \eqref{eq:psitetrahedron}, but, again, this is a tedious calculus and the final formula remains a complicated analytical expression \citep[see also][]{c12}. If we replace the pyramidal bump by point mass (or a sphere) at $\lambda/4$ from the base (the center of mass), then the potential at vertex V$^{\rm new}_{ijk}$ is $-4 G m/3\lambda$, where $m=\frac{4}{3} \rho a_1 a_2 \lambda$ is the mass of the pyramid. As the self-gravitating potential of a pyramidal bump generated by stretching V$_{ijk}$, we therefore retain 
\begin{flalign}
  \delta \Psi_{\rm self} \sim -\frac{16}{9} G \rho a_1 a_2 
  \label{eq:devpsi2}
\end{flalign}
We can even link $a_1 a_2$ to the number $T$ of triangles by a supplementary hypothesis. If the surface of the body is triangulated in a rather uniform manner (there are no large and tiny triangles) and has total area ${\cal A}$, then $a_1 a_2 \simeq 2 {\cal A}/T$. So, by dividing Eqs. \eqref{eq:devpsi} and \eqref{eq:devpsi2} by $G \rho L^2$ and setting $\Delta \Psi \approx \delta \Psi_{\rm self}$, which is the typical absolute error in potential values, leads to
\begin{flalign}
  \frac{32}{9} \frac{{\cal A}}{L^2 T} \simeq \tilde{g} \frac{\lambda}{L},
\end{flalign}
 where $\tilde{g}$ is the magnitude of gravitational acceleration at the surface in units of $G \rho L$. By solving this expression for the number of triangles, we find
\begin{flalign}
 T \simeq \frac{32}{9} \frac{{\cal A}}{L^2} \times \frac{L}{\lambda}\times \frac{1}{\tilde{g}} \equiv T_\lambda
  \label{eq:criticalt}
\end{flalign}
This formula gives the resolution, in term of number of triangles, beyond which a shape error has the same impact in the potential as a bump or a cavity. Except for very elongated bodies, $\tilde{g}$ is of the order of unity, and so Eq. \eqref{eq:criticalt} is $T_\lambda \simeq 32 {\cal A}/9 L \lambda$. Note that, if the body is close to spherical, then ${\cal A} \approx 4 \pi L^2$, and so Eq. \eqref{eq:criticalt} simplifies for
\begin{flalign}
 T_\lambda \simeq \frac{128 \pi}{9}  \frac{L}{\lambda},
\end{flalign}
hence the formula in the abstract.

 \begin{table}[h]
   \centering
   \caption{Dimensionless potential at the central vertex V$_1$ of plate $f=1$ when its altitude $z_1$ is changed for $z_1 \pm \lambda$ (columns 5 and 6), compared to the unperturbed surface (colum 3).}
   \begin{tabular}{rrrrrl}
      &           &                  \multicolumn{3}{c}{$\Psi/GL^2$ at the surface}\\
  $q$ & $T$       & at V$_1$ & at V$^{\rm new}_1(z_1+\lambda)$ & V$^{\rm new}_1(z_1-\lambda)$ & Comment \\  \hline
$512$ & $3145728$ & $\underline{-0.9777}9130$ & $\underline{-0.977}20367$ & $\underline{-0.9783}7893$ & Reference $\Psi_{\rm ref}$\\
$256$ & $786432$  & $\underline{-0.9777}7931$ & $\underline{-0.977}19272$ & $\underline{-0.9783}6589$ \\
$128$ & $196608$  & $\underline{-0.9777}3042$ & $\underline{-0.977}14598$ & $\underline{-0.9783}1487$ \\
$64$  & $49152$   & $\underline{-0.977}50019$ & $\underline{-0.97}692036$ & $\underline{-0.978}08005$  \\
$32$  & $12288$   & $\underline{-0.97}681906$ & $\underline{-0.97}624908$ & $\underline{-0.97}738906$\\
$16$ & $3072$     & $\underline{-0.97}378461$ & $\underline{-0.97}323698$ & $\underline{-0.97}433227$\\ \hline
   \end{tabular}
  \label{tab:eros_psiatv0s}
 \end{table}

\subsection{Application to 433 Eros}
  
In the case of 433 Eros, we have $\lambda/L \approx 0.00034$ with $\lambda=6$ m, and then we get $\Delta \Psi \approx 1.684 \times \lambda/L \approx 5.7 \times 10^{-4}$ from Eqs. \eqref{eq:psigmean} and \eqref{eq:devpsi}. This implies that the error level can not allow to fix $\Psi$ with more than $3$ to $4$ significant digits at most, whatever the resolution. As an illustration, we have replaced the altitude $z_1$ of the central vertex of plate $f=1$ by $z_1 \pm \lambda$, and we have computed the potential at the two new positions V$^{\rm new}_{ijk}$. Columns 4 and 5 of Table~\ref{tab:eros_psiatv0s} list the results. We find $\Psi($V$_1)/GL^2 \approx -0.97779 \pm 0.0006$ and we see that no more than $3$ digits are fixed for $q \ge 64$. This is coherent with Eqs. \eqref{eq:psigmean} and \eqref{eq:devpsi}.

\begin{figure}[h]
       \centering
       \includegraphics[width=8.9cm,trim={3cm 2.5cm 2.2cm 2cm},clip,angle=0]{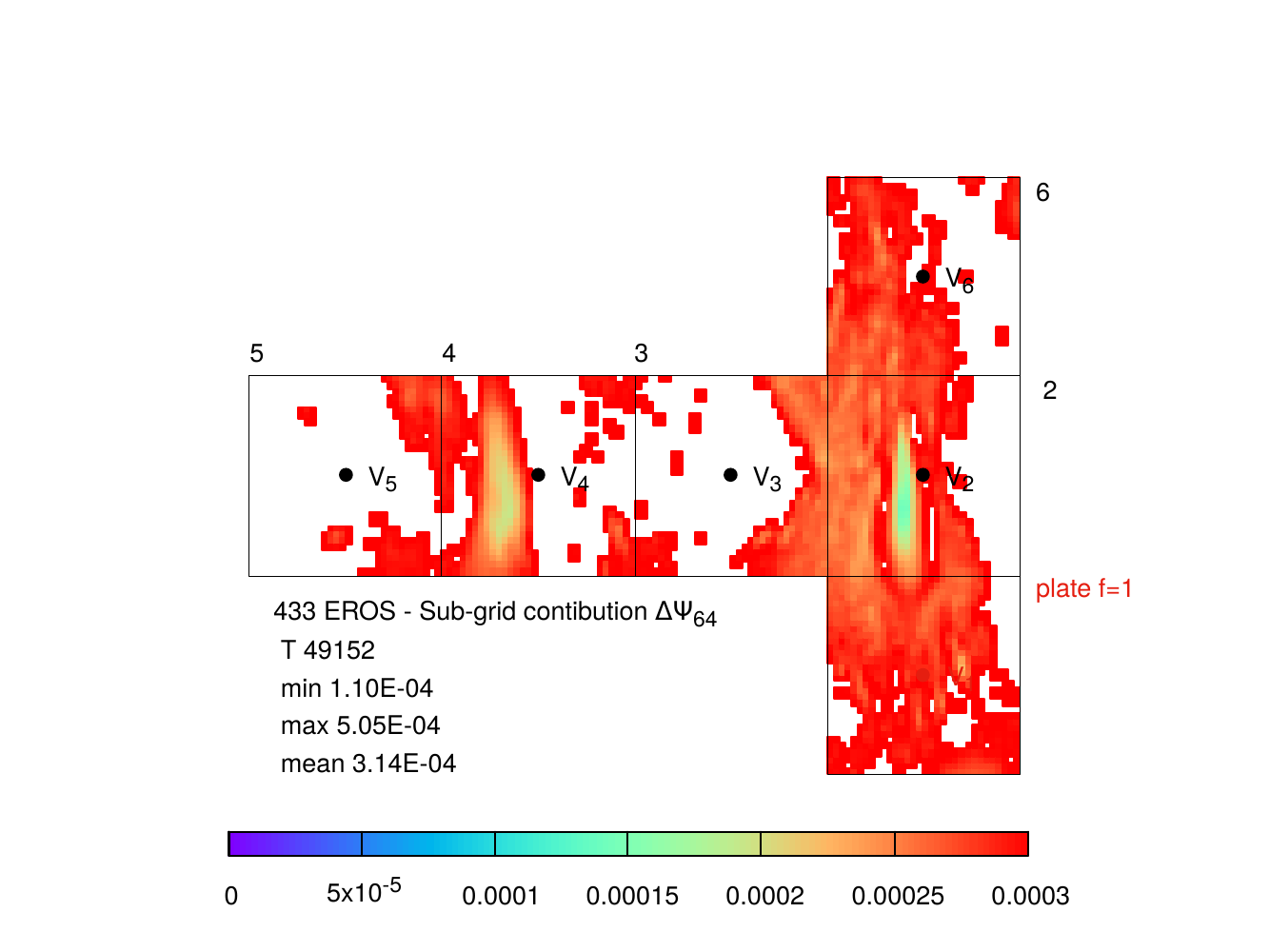}\\
       \includegraphics[width=8.9cm,trim={3cm 0.cm 2.2cm 3cm},clip,angle=0]{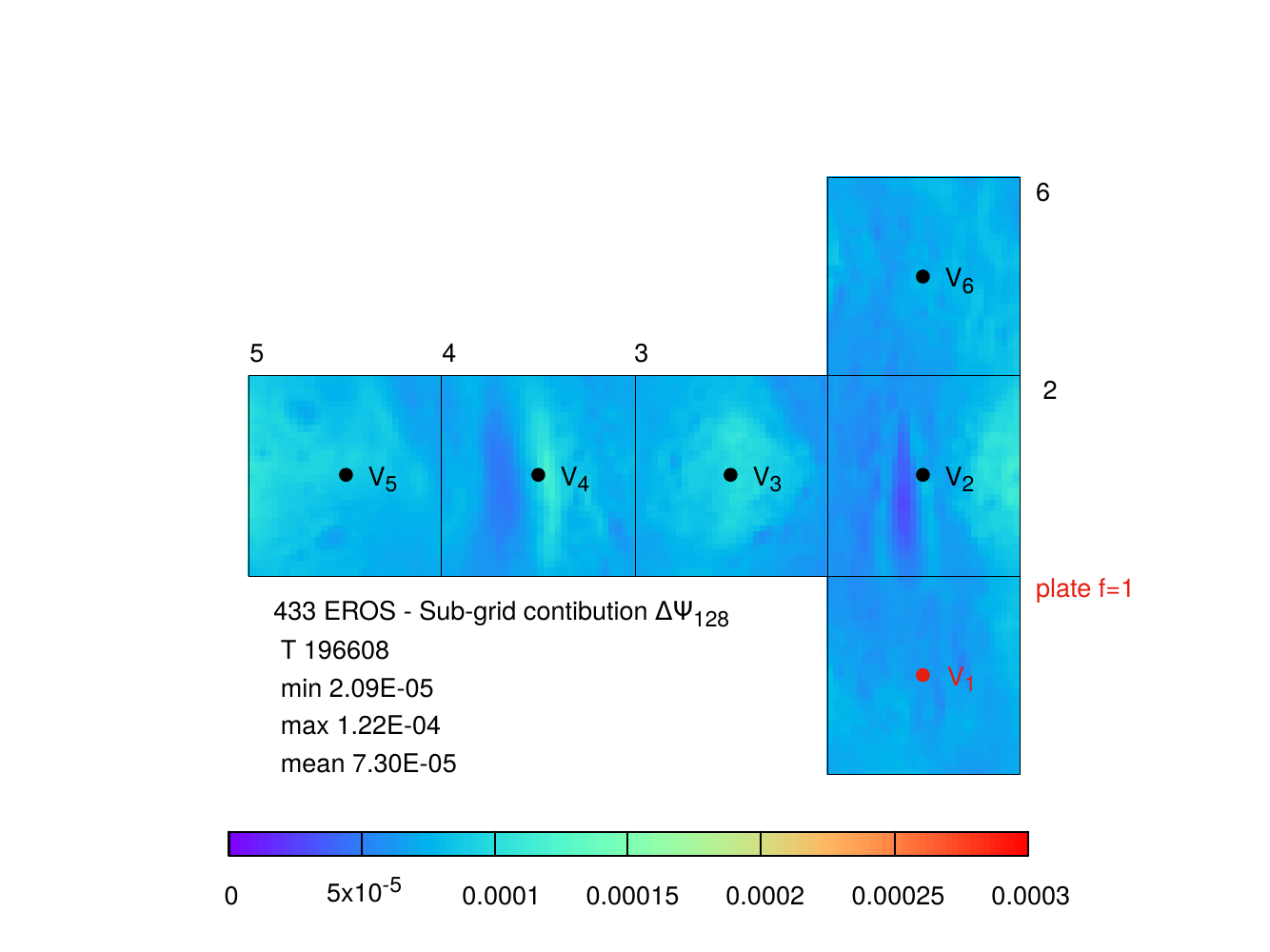}\\
       \caption{Deviation $\Delta \Psi_q$ showing the contribution of cavity and bumps present below a given resolution, $q=64$ ({\it top}) and $q=128$ ({\it bottom})}
       \label{fig:eros_deltapsismooth.pdf}
\end{figure}

\begin{figure}[h]
       \centering
       \includegraphics[height=4.5cm,trim={5.5cm 3.5cm 0.cm 4.5cm},clip]{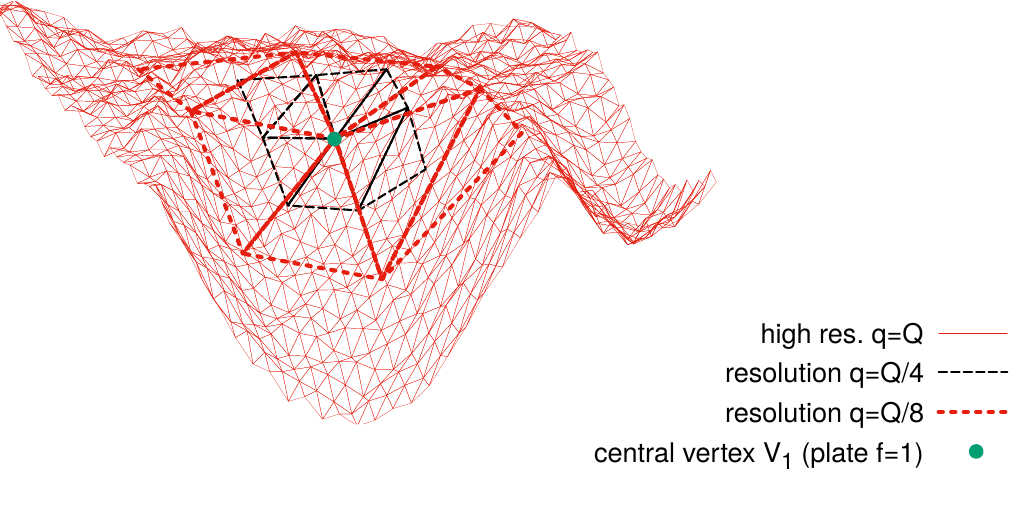}
       \caption{Same caption as for Fig.~\ref{fig:eros_smooth.pdf} by after moving all vertices randomly, with a maximum shift of $\lambda = 6$ m}
       \label{fig:eros_err.pdf}
\end{figure}


\begin{figure}
       \centering
       \includegraphics[width=8.9cm,trim={3cm 2.5cm 2.2cm 2cm},clip,angle=0]{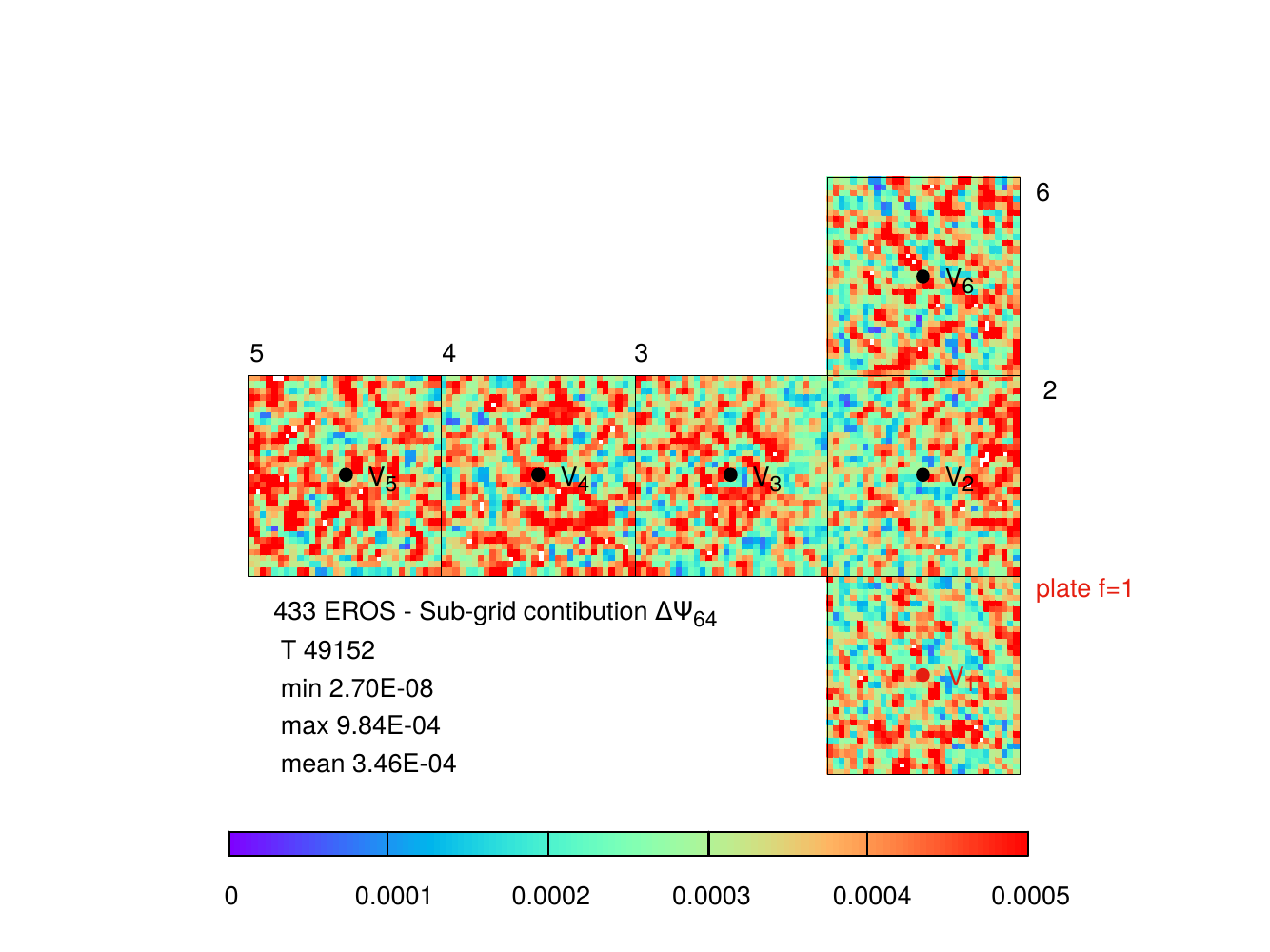}\\
       \includegraphics[width=8.9cm,trim={3cm 0.cm 2.2cm 3cm},clip,angle=0]{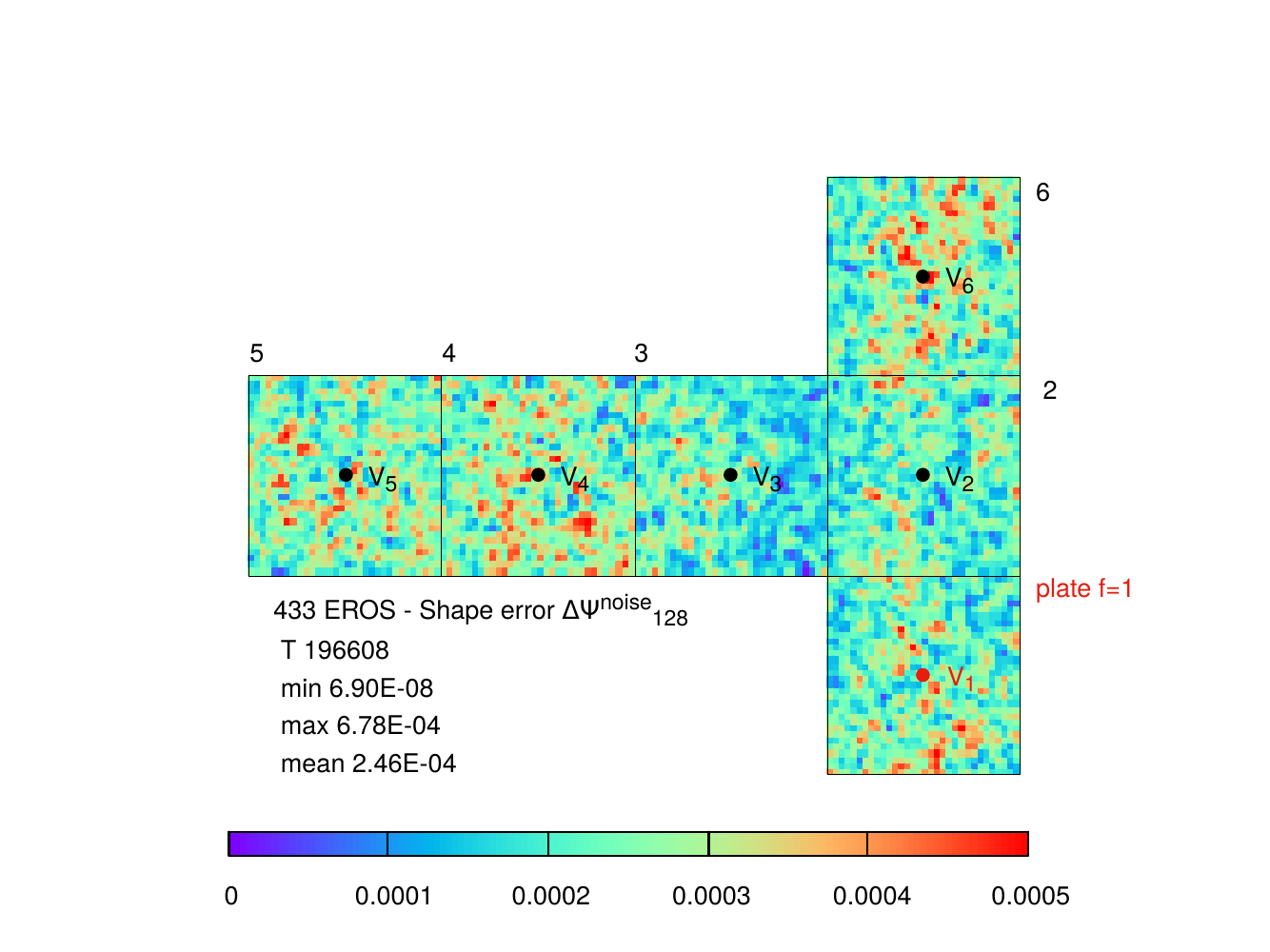}
        \caption{Same caption as for Fig.~\ref{fig:eros_deltapsierr.pdf} but for $\Delta \Psi^{\rm noise}_q$}
       \label{fig:eros_deltapsierr.pdf}
\end{figure}

We can test the prescription established in Sect.~\ref{sec:prescription}. Feeding Eq. \eqref{eq:criticalt} with appropriate values for 433 Eros gives $T_\lambda = 22889$, which corresponds to $q_\lambda \approx 44$, in between grid parameters $q=32$ and $q=64$. It means that, with Eros data, the resolution corresponding to $q=32$ (i.e., about $0.5$ km on average) is in principle sufficient to reproduce the bulk of the potential (and acceleration), with $3$ significant digits. Equivalently, for $q \gtrsim 64$ (i.e., below $260$ m on average), cavities and bumps present at smaller scales leave imprints comparable to altimetric uncertainties. To prove this point, we need two kinds of experiments:
  \begin{itemize}
  \item quantifying the contribution of sub-grid irregularities. This is performed by comparing the potential $\Psi_q$ obtained with grid parameter $q$ to the reference $\Psi_{\rm ref} \equiv \Psi_Q$. We define $\Psi_q-\Psi_{\rm ref} = \Delta \Psi_q$, which gives the contribution of sub-grid irregularities. The results are displayed in Fig. \ref{fig:eros_deltapsismooth.pdf} for $q=Q/8=64$ and for $q=Q/4=128$ (these correspond to resolutions of $260$ m and $131$ m at the surface, respectively; see Table~\ref{tab:erostijdata}). Wee see that the contribution of sub-grid irregularities, in units of $G \rho L^2$, is about $0.0003$ at $q=64$ and about a quater this value at $q=128$. This is similar to what is observed for vertex V$_1$.
  \item estimating the contribution of altimetric uncertainties. In this purpose, we have shifted all vertices of the finest grid according to
\begin{equation}
  \mathbf{OV}_{ijk}^{\rm noise}= \mathbf{OV}_{ijk}\times \left[1 + (-1+2 \sigma )\frac{\lambda}{\text{OV}_{ijk}}\right],
  \label{eq:eroswithnoise}
  \end{equation}
where $\sigma \in [0,1]$ is a random number, and $\lambda=6$ m (in this way, the V$^{\rm noise}_{ijk}$ can be below or above the surface). A similar technique has been used in \cite{bpm20}. Figure~\ref{fig:eros_err.pdf} shows the ``noisy'' surface  around the central vertex $V_1$ so obtained. The new potential, $\Psi_q^{\rm noise}$, is then compared to the reference corresponding to the unperturbed surface. The deviation is now denoted $\Psi_q^{\rm noise}-\Psi_{\rm ref}= \Delta \Psi^{\rm noise}_q$. The results are displayed in Fig. \ref{fig:eros_deltapsierr.pdf}, again for $q=Q/8=64$ and for  $q=Q/4=128$. We notice that the contribution of errors in the shape of the asteroid are of the order of $\lambda/L$ on average, and this level remains the same if we consider the next grid parameter $q=256$.
  \end{itemize}

  If we now compare Figs.~\ref{fig:eros_deltapsismooth.pdf} and \ref{fig:eros_deltapsierr.pdf}, we clearly see that $\Delta \Psi_{64} \gtrsim \Delta \Psi^{\rm noise}_{64}$ while $\Delta \Psi_{128} \lesssim \Delta \Psi^{\rm noise}_{128}$. This proves that the contribution of altimetric uncertainties dominates over sub-grid irregularities for $q \gtrsim 64$, in agreement with the prescription.

\section{Comments on the computing time and error reduction}
\label{sec:gain}

If $\tau(\ell)$ is the typical computing time at mesh level $\ell$ to get the potential at all nodes/vertices, then we have for the RRE technique
\begin{flalign}
  \tau_{\text{RRE}}(\ell) = \sum_{l=0,\ell}{\tau(l)} & \propto \sum_{l=0,\ell}{\left(2^l\right)^4}\\ & \propto 16^\ell \times \left(1+ \frac{1}{16}+\frac{1}{256}+...\right)  \simeq 1.07 \, \tau(\ell),
  \nonumber
\end{flalign}
where $\ell$ remains relatively small\footnote{For $\ell \rightarrow \infty$, the sum tends to $\pi^4/90 \approx 1.082$}. In terms of computing time, the extra cost represented by the acceleration therefore appears marginal, when compared to the number of digit fixed. In case of a full parallel treatment, we have $\tau(k) \propto 2^{2k}$ and then $\tau_{\text{RRE}}(\ell)  \simeq 1.33 \, \tau(\ell)$, which is still very satisfactory.\footnote{For $\ell \rightarrow \infty$, that sum tends to $\pi^2/6 \approx 1.64$}

Beyond this resolution, all the details of the terrain and shape errors contribute at the same level, and can be regarded as a background noise.

 \begin{table}[h]
   \centering
   \caption{Same caption as for Tab.le~\ref{tab:slopes}, but after applying the Repeated Richardson Extrapolation (RRE) in columns 2, 4 and 6; see also Figs.~\ref{fig:test_RRE_sphere.pdf} and \ref{fig:test_RRE_dumbell.pdf}}
   \begin{tabular}{llllllll}
            & \multicolumn{2}{c}{Inside ($k=1$)} & \multicolumn{2}{c}{Surface ($k=2$)} &  \multicolumn{2}{c}{oOutside ($k=3$)} \\\hline
            & $\frac{2\alpha}{\alpha_{\rm RRE}}$ & $\frac{4\alpha}{\alpha_{\rm RRE}}$ & $\frac{2\alpha}{\alpha_{\rm RRE}}$ & $\frac{4\alpha}{\alpha_{\rm RRE}}$ & $\frac{2\alpha}{\alpha_{\rm RRE}}$ & $\frac{4\alpha}{\alpha_{\rm RRE}}$ \\\hline
Sphere      & $0.472$ & $0.945$ & $0.926$ & $1.851$ & $0.540$ & $1.080$ \\
Spheroid    & $0.649$ & $1.299$ & $0.870$ & $1.741$ & $0.629$ & $1.258$ \\
Triaxial    & $0.607$ & $1.213$ & $0.873$ & $1.747$ & $0.596$ & $1.192$ \\
Dumbell     & $0.529$ & $1.058$ & $0.910$ & $1.820$ & $0.577$ & $1.154$\\
Lemon-shape & $0.596$ & $1.192$ & $1.103$ & $2.207$ & $0.631$ & $1.262$ \\
Torus       & $0.623$ & $1.246$ & $0.990$ & $1.981$ & $0.608$ & $1.217$\\\hline
    \end{tabular}
  \label{tab:gains}
  . The statistics are performed by omitting the first point and points beyond the knee, i.e. for $32 \le N \le 256$. The numbers in columns 3, 5 and 7 ({\it bold}) represent the gain per decade in $N$ with respect to the direct method, i.e. $\alpha - \alpha_{\rm RRE} \equiv \Delta \alpha$
 \end{table}
 
Regarding accuracy, the gain obtained at a given resolution is simply measured by the difference $E-E_{\text{RRE}} \propto \Delta \alpha$. Table~\ref{tab:slopes_RRE} contains (columns 3, 5 and 7) the extra number of digits fixed  per decade in $N$. Finally, it is interesting to couple the accuracy and computing time. Let us consider a given error level, reached with a grid $N \times M$ by direct summation and with $N_{\rm RRE} \times M_{\rm RRE}$ by applying the RRE technique. Equating $E(N)$ and $E_{\rm RRE}(N_{\rm RRE})$ leads to the relationship
\begin{flalign}
  N_{\rm RRE} = e^{(\beta_{\rm RRE}-\beta)/\alpha_{\rm RRE}}N^{\alpha/\alpha_{\rm RRE}}
\end{flalign}
and then we get
\begin{flalign}
  \tau_{\rm RRE}  \propto
  \begin{cases}
 N^{4\alpha/\alpha_{\rm RRE}} & \text{per grid (sequential processing)},\\
 N^{2\alpha/\alpha_{\rm RRE}} & \text{per vertex (parallel processing)}.
  \end{cases}
\end{flalign}
The exponents $2\alpha/\alpha_{\rm RRE}$ and $4\alpha/\alpha_{\rm RRE}$ are listed in Table~\ref{tab:gains} for the six solid bodies considered in Sect.~\ref{sec:solution}.

\section{Discussion}
\label{sec:discussion}

In potential theory, there are a few finite size, solid bodies that admit a closed form expression for the gravitational potential and acceleration \citep{durand64}. The tetrahdron is one of these (see the references in the introduction). As a consequence, for any system whose volume (surface) can be decomposed into a series of tetrahedra (resp. triangles), the solution to the Poisson equation is analytical. If the process of surface triangulation can incorporate nested grids as assumed here (with $N \propto 2^\ell$ nodes per direction), then we can take benefit of classical extrapolation methods to accelerate the convergence of potential and gravity toward values that would be obtained at infinite resolution. Note that the method applies to other quantities, like the mass, surfce area, moment of inertia, etc., but imposes that the surface in question is smooth enough. A few simple shapes have been tested, and it happens that the RRE technique is very efficient for points located outside and inside matter; see Table~\ref{tab:slopes_RRE}. This is particularly interesting for following test-particles orbiting around, which requires a very large number of evaluations. At the surface of body, the convergence rate is still excellent, although significantly lower (see the table, again). Surface values are fundamental to study the mechanics of the terrain itself (tectonics, crumbling, piles, etc.). A possible extension of the work concerns the mesh. In Sect.~\ref{sec:solution}, surface triangulation is based on spherical coordinates, which leaves triangles with inequal shapes and surface area, in between the poles and the equator. In some sense, there is possible bias, not in the approach, but in the tests we have performed. For instance, testing the Fibonacci lattice on a sphere would be interesting.

As discussed in Sect.~\ref{sec:eros}, real body like asteroids, comets, dwarf planets have irregular and fractal terrains, inherent in the mechanisms of formation and evolution. In such a case, the RRE technique does not offer a great gain, with respect to perfect shapes. This has been illustrated with 433 Eros. However, we have shown that the errors in the shape of the asteroid, which are of the order of a few meters, leaves imprints in potential values at the surface comparable to and even larger than terrain irregularities at resolutions of the order of $261$ m and below, corresponding to the grid parameter $q \gtrsim 64$ \cite[see also][]{bpm20}. This limit is expressed by Eq. \eqref{eq:criticalt} and it varies as the inverse of the typical topographic uncertainties associated to a given shape model. It is established in very simple conditions and therefore, it remains an indicator. The criterion must therefore be used with caution, but it applies to most shapes. This is especially interesting for dynamical studies, as it permits to select an optimal spatial resolution for the gravity source. The paper, which focuses on the gravitational potential, could be continued by analyzing the gravitational acceleration and gravitational slopes \cite[e.g.,][]{PROCKTER2002,THOMAS2002,CHAPMAN2002}, which are fundamental regarding the mechanical stability of the terrains (formation of ponds, landslides and avalanches). 
 
\bmhead{Acknowledgements}
It is a pleasure to thank E.E. Palmer at Planetary Science Institute for his help in using the data of asteroid 433 Eros and advices about uncertainties, L. Jorda at the Laboratoire d'Astrophysique de Marseille and C. Staelen at the Institut de Physique du Globe de Paris for fruitful inputs on the manucript before submission. I would like to thank the anonymous referee for very interesting comments and criticisms on the method, and several suggestions to improve the reading of the article.



\appendix

\section{Formula for $\delta \Psi_{ijk}$}
\label{app:i}

If we set $\Delta(a,b;x)=\sqrt{(1+a^2 ){x}^2+b^2}$, then
\begin{flalign}
  I(a,b;h) = \int_{0}^{x} {\atanh \frac{a x'}{\Delta(a,b;x')} dx' }& = x \atanh \frac{a x}{\Delta(a,b;x)}\\
  & \quad -b \arctan \frac{\Delta(a,b;x)}{ab} + |b| \arctan \frac{1}{a}, \nonumber
\end{flalign}
which is the formula for Eq. \eqref{eq:i}. If $b=0$, then only the first term in the right-hand-side is present.

\section{Location of vertex A in the reference coordinate system}
\label{app:avertex}

Using a Cartesian reference $($O$x$,O$y$,O$z)$ with basis $(\ex,\ey,\ez)$, we have to determine the equation of the triangle's plane. Note that the transformation matrix $(\ex,\ey,\ez) \leftrightarrow (\exprim,\eyprim,\ezprim)$ is already known as soon as any triangle  $\text{triangle T}_i\text{V}_j\text{V}_k$ is identified. Let $\ezprim=u\ex+v\ey+w\ez$  and P$(x,y,z)$ in the same in the reference coordinate system. The coordinates of point A are found by imposing simultaneously $\mathbf{AP} \propto \mathbf{e}_z$ and $\mathbf{AV}_i \cdot \mathbf{e}_z=0$. If $u \ne 0$, then we have
\begin{flalign}
  \begin{cases}
    x_A=x+\delta x,\\
y_A=y+\frac{v}{u}\delta x,\\
z_A=z+\frac{w}{u}\delta x,
  \end{cases}
  \label{eq:pointa}
\end{flalign}
where $\delta x = -u\left[(x-x_i)u+(y-y_i)v+(z-z_i)w\right]$ and $(x_i,y_i,z_i)$ are the Cartesian coordinates of vertex V$_i$. If $u=0$, then another formula must be employed (see the Appendix \ref{app:avertex}). If $v \ne 0$, then
\begin{flalign}
  \begin{cases}
    x_A=x+\frac{u}{v}\delta y,\\
y_A=y+\delta y,\\
z_A=z+\frac{w}{v}\delta y,
  \end{cases}
\end{flalign}
  where $\delta y = -v\left[(x-x_i)u+(y-y_i)v+(z-z_i)w\right]$. Optionally, if $w \ne 0$, then
\begin{flalign}
  \begin{cases}
    x_A=x+\frac{u}{w}\delta z,\\
y_A=y+\frac{v}{w}\delta z,\\
z_A=z+\delta z,
  \end{cases}
\end{flalign}
where $\delta z = -w\left[(x-x_i)u+(y-y_i)v+(z-z_i)w\right]$.

\section{Comments}
\label{app:comments}
           
\noindent
    {\bf Potential at a vertex}. According to Eq. \eqref{eq:deltapsiijk}, when the point P belongs to the same plane as the triangle ${\cal T}_{ijk}$, whatever the distance to it, then $Z'_{ijk}=0$. It means that, in such a case, the triangle does not contribute in the summation, i.e., $\delta \Psi_{ijk}=0$. This is still true if P coincides with one of the $3$ vertices. But if P is a vertex for ${\cal T}_{ijk}$, it is also a vertex for a few neighboring triangles, implying that the contribution $\delta \Psi$ of all triangles sharing the same vertex in question is zero. With a spherical mesh, at the exception of the two poles, each vertex belongs to six triangles, and at least six contributions $\delta \Psi$ are strictly zero in this case.\\

\noindent
{\bf Case of a tetrahedron}. If the surface ${\cal S}$ is just a tetrahedron, then there are $T=4$ triangles in total, and four vertices V$_i(\mathbf{r}_i)$ with $i \in [1,4]$. Consequently to the point quoted above, the gravitational potential at vertex V$_1$ of a tetrahedron is given by a single contribution
\begin{flalign}
  \Psi(\mathbf{r}_1) = \delta \Psi_{234}(\mathbf{r}_1).
  \label{eq:psitetrahedron}
\end{flalign}

\noindent
{\bf Gravitational acceleration}. The knowledge of the acceleration $\mathbf{g}(\mathbf r)=-\nabla \Psi(\mathbf r)$ is fundamental for many applications, in particular for dynamical studies. As soon as the individual potentials $\delta \Psi_{ijk}$ are calculated and stored, no additional effort is needed to compute $\mathbf{g}$. {\bf Each} contribution $\delta \mathbf{g}_{ijk} =-\nabla \delta \Psi_{ijk}$ is colinear with the local normal vector $\mathbf{n}_{ijk}$ and involves Eq.\eqref{eq:integijk}; see the Appendix \ref{app:acceleration} for the derivation. It turns out that we get an exact expression for the acceleration \citep[e.g.][]{hk96}, and we can avoid numerical derivatives which are always sources of errors.
           
\section{Gravitational acceleration}
\label{app:acceleration}

The gravitational acceleration is $\mathbf{g}(\mathbf r)=-\nabla \Psi(\mathbf r)$. As we have
\begin{flalign}
\frac{x-x'}{|\mathbf{r}-\mathbf{r'}|^3} = \nabla \cdot \frac{\mathbf{e}_x}{|\mathbf{r}-\mathbf{r'}|},
\end{flalign}
we find from Eq. \eqref{eq:tripleint}. for the first component
\begin{flalign}
  - \partial_x \Psi(\mathbf r) = -G \rho \iint{\frac{\mathbf{e}_x}{|\mathbf{r}-\mathbf{r'}|}\cdot \mathbf{n}(\mathbf{r'})dS'}
  \label{eq:tripleint_accx}
\end{flalign}
For a triangle ${\cal T}_{ijk}$, the contribution is
\begin{flalign}
- \partial_x \delta \Psi_{ijk}(\mathbf r) = -G \rho \iint_{\text{triangle } {\cal T}_{ijk}}{\frac{\mathbf{e}_x}{|\mathbf{r}-\mathbf{r'}|}\cdot \mathbf{n}(\mathbf{r'})dS'}
\end{flalign}
where $\mathbf{n}(\mathbf{r'})$ is constant for any point P' inside ${\cal T}_{ijk}$. This expression is therefore given by
\begin{flalign}
- \partial_x \delta \Psi_{ijk}(\mathbf r) = -G \rho n_x \iint_{\text{triangle } {\cal T}_{ijk}}{\frac{dS'}{|\mathbf{r}-\mathbf{r'}|}}
\end{flalign}
where $n_x=\mathbf{n} \cdot \mathbf{e}_x$. We see that this integral as already been calculated, and it can therefore been imployed to deduce the components of the gravitational acceleration with no additional effort, namely
\begin{flalign}
- \nabla \delta \Psi_{ijk}(\mathbf r)&= -G\rho \mathbf{n}_{ijk} \iint_{\text{triangle } {\cal T}_{ijk}}{\frac{dS'}{|\mathbf{r}-\mathbf{r'}|}},\\
  &= -G\rho \mathbf{n}_{ijk} \left[ s_{ij} I_{\text{AV}_i\text{V}_j} +   s_{jk} I_{\text{AV}_j\text{V}_k} +  s_{ki} I_{\text{AV}_k\text{V}_i}\right]
\label{eq:tripleint_acc}
\end{flalign}
By using Eq. \eqref{eq:deltapsiijk}, we have
\begin{flalign}
  \delta \mathbf{g}_{ijk}(\mathbf r)= -2G\rho \mathbf{n}_{ijk} \frac{\delta \Psi_{ijk}(\mathbf r)}{Z}
\end{flalign}
and so
\begin{flalign}
  \mathbf{g}( \mathbf r)&= \sum_{\text{all triangles}}{\delta \mathbf{g}_{ijk}}\\
  &=-2G\rho \sum_{\text{all triangles}}\mathbf{n}_{ijk} \frac{\delta \Psi_{ijk}(\mathbf r)}{Z_{ijk}}.
  \nonumber
\end{flalign}

\section{Convergence acceleration by using the Repeated Richardson Extrapolation (RRE)}
\label{app:rre}

Richardson's Extrapolation (RE) is a standard numerical technique for improving quadratures and derivatives. It uses two evaluations of a given quantity $f$ (supposed regular enough) obtained at two different resolutions $h$ and $h'<h$, namely $f_h$ and $f_{h'}$, and assumes that the absolute error has leading term $h^n$, where $n$ is the order of the error. In these conditions, a better estimate $f^*$ is given by the formula
\begin{equation}
f^* \approx f_{h'} + \frac{f_{h'}-f_h}{ \left(\frac{h}{h'}\right)^n-1}.
\end{equation}
Usually, the setting is $2h'=h$, which corresponds to two computational grids with same boundaries but with, respectively, $2^{l}+1$ and $2^{l+1}+1$ equally space nodes, where $l \ge 0$ is the discretization level (i.e., the coarser grid is contained in the finer one). In fact, the process can be repeated by considering that the error is a power series in $h$, which assumes that the function is ``regular'' enough \citep[e.g.,][]{opac-b1132370}. In particular, if the error is second-order in the mesh spacing, then the series contains only odd powers of $h$. Let $A_{l,0} \equiv f_{h_l}$ denote a series of approximations for $f$ obtained at various resolutions $2h_{l+1}=h_l$, with $l \ge 0$. Based upon this series, we then form a triangular table where the elements $A_{l,k}$ are found from the formula
\begin{equation}
A_{l,m} = A_{l,m-1} + \frac{A_{l,m-1}-A_{l-1,m-1}}{4^m-1},
\label{eq:richardsonextrapolationtable}
\end{equation}
for $m \in [1,l]$ and $l\ge 1$. The last element $A_{l,l}$ is the improved value of $f$. The error is estimated to be of the order of $h^{2(l+1)}$. This is the technique of Repeated Richardson Extrapolation (RRE). In the present paper, $l \equiv \ell \ge 3$, then $m \ge 3$. Accordingly,  Eq. \eqref{eq:richardsonextrapolationtable} is to be used for $m \ge 4$, and the denominator $4^m-1$ must be replaced by $4^{m-3}-1$.

\section{Example at the center of the sphere (con't)}
\label{app:qple}

\begin{table}[h]
  \caption{Same caption as for tab. \ref{tab:RREatthecenter_qp} but in quadruple precision.}
  \begin{tabular}{llllll}
$\ell$ & $A_{\ell,3}$            & $A_{\ell,4}$             & $A_{\ell,5}$            & $A_{\ell,6}$    \\ \hline
 3& $-5.2658724606271720$ &\\
 4& $\underline{-6.}0177824547115461$ & $\underline{-6.2}684191194063375$ \\
 5& $\underline{-6.2}161323588550541$ & $\underline{-6.28}22489935695567$ & $\underline{-6.2831}709851804380$ \\
 6& $\underline{-6.2}663780221957858$ & $\underline{-6.2831}265766426963$ & $\underline{-6.283185}0821809056$ & $\underline{-6.28318530}59428178$ \\
 7& $\underline{-6.2}789807304634097$ & $\underline{-6.28318}16332192843$ & $\underline{-6.28318530}36577235$ & $\underline{-6.28318530717}32286$ \\
 8& $\underline{-6.28}21339907449427$ & $\underline{-6.283185}0775054538$ & $\underline{-6.2831853071}245317$ & $\underline{-6.2831853071795}604$ \\
 9& $\underline{-6.28}29224673043457$ & $\underline{-6.283185}2928241466$ & $\underline{-6.28318530717}87261$ & $\underline{-6.2831853071795864}$ \\
 10& $\underline{-6.2831}195965378556$ & $\underline{-6.28318530}62823589$ & $\underline{-6.2831853071795}730$ & $\underline{-6.283185307179586}5$ \\ \\
$\ell$ & $A_{\ell,7}$             & $A_{\ell,8}$            & $A_{\ell,9}$             & $A_{\ell,10}$\\ \hline
 3&  \\
 4&  \\
 5&  \\
 6&  \\
 7&  $\underline{-6.283185307178}0537$ \\
 8&  $\underline{-6.28318530717958}53$ & $\underline{-6.283185307179586}8$ \\
 9&  $\underline{-6.2831853071795865}$ & $\underline{-6.2831853071795865}$ & $\underline{-6.2831853071795865}$\\
 10& $\underline{-6.2831853071795865}$ & $\underline{-6.2831853071795865}$ & $\underline{-6.2831853071795865}$ & $\underline{-6.2831853071795865}$ \\  \\
 &&& reference (exact)  &   $-6.2831853071795864$ \\
    \end{tabular}
  \label{tab:RREatthecenter_qp}
\end{table}

\end{document}